# Stochastic partial differential equation model for environmental DNA dynamics in river environments

Hidekazu Yoshioka[1,*]

[1] Graduate School of Advanced Science and Technology, Japan Advanced Institute of Science and Technology, 1-1 Asahidai, Nomi, Ishikawa, Japan

* E-mail: yoshih@jaist.ac.jp, ORCID: https://orcid.org/0000-0002-5293-3246

**Abstract**

Environmental DNA (eDNA) has emerged as a novel tool for quantifying the seasonal abundance of aquatic species in water bodies; however, its mathematical modeling is still at a germinating stage because of its mechanistic uncertainties. We propose a first-step mathematical and computational framework for the eDNA dynamics of migratory fish based on a novel stochastic partial differential equation model with a delayed source input. The model governs spatiotemporal eDNA concentration in rivers where the source comes from a stochastic differential equation for the migration dynamics of the fish. The affine nature of the model facilitates its theoretical analysis, including the guarantee of well-posedness and the closed-form derivation of the Laplace functional despite the proportionality coefficient of the multiplicative noise term being non-Lipschitz. We also propose a discretization scheme for the model that theoretically generates nonnegative numerical solutions. We finally apply the proposed model to eDNA concentration data sampled from midstream reaches of a river system and perform sensitivity analysis.

***Statements & declarations***

**Acknowledgments:** We would like to express our sincere gratitude to the members of the Hii Rivet Fisheries Cooperative for their tremendous cooperation during the field survey. The eDNA analysis was outsourced to the Environmental Pollution Research Center Co., Ltd.

**Funding:** This study was supported by the Japan Science and Technology Agency (JST): PRESTO No. JPMJPR24KE.

**Conflict of Interests:** The authors have no relevant financial or nonfinancial interests to disclose.

**Data Availability:** The data will be made available upon reasonable request to the corresponding author.

**Declaration of Generative AI in Scientific Writing:** The author used Rubriq provided by American Journal Experts to proofread the manuscript.

# 1. Introduction

## 1.1 Background

### 1.1.1 Environmental DNA and fish migration

Analyzing the abundance of aquatic organisms in water bodies, such as rivers, lakes, and seas, is crucial across diverse academic fields, including ecology, biology, and fisheries science. The abundance of aquatic organisms can be measured directly or indirectly. Direct methods, such as capture surveys [1] and passive integrated transponder tagging [2], are commonly used; however, they not only require a vast amount of labor costs but also unnecessarily disrupt the ecosystem and possibly apply some surgical procedures to sampled individuals [3].

Environmental DNA (eDNA) analysis has recently emerged as a novel tool for indirectly detecting the presence/absence and quantifying the abundance of aquatic organisms [4-6]. This is a technique for counting the DNA copy number of a target species from a water sample at a location in a water body, and therefore, its influence on aquatic environments and ecosystems is minimal. eDNA has been applied to a variety of aquatic species and water bodies, including but not limited to the diversity assessment of aquatic species in seas [7] and river systems [8], biomass and density estimation [9,10], and blue carbon assessment [11].

Despite the recent significant progress in understanding the eDNA of aquatic organisms, mathematical modeling to consistently connect population dynamics, shedding from individuals and their decay, and transport in water bodies has not been established [12]; however, there is an accumulation of knowledge that potentially serves as hints for the modeling of the elementary processes. Linear regression approaches between hydrological variables and eDNA concentrations have been investigated in rivers [13,14]. Linear ordinary differential equation (ODE) models for eDNA shedding have been studied [15,16], and sensitivity analyses against water temperature have been conducted [17,18]. Spatial eDNA decay has also been studied based on linear ODE models [19,20]. Nonlinear ODE models that govern population dynamics and eDNA shedding in a water column have been formulated along with a theoretical species detection study [21]. Lagrangian transport models of eDNA have been examined along with uncertainty analysis [22,23]. A graph-based approach for the transport of eDNA in river systems was presented by Carraro et al. [24] and has later been applied to modeling efficient sampling strategies [25]. An artificial intelligence approach for the relationships among eDNA and multiple water quality indices has been applied to a large river [26].

As outlined above, while there have been advances in field data and the modeling of elementary processes related to eDNA, studies linking these processes to fish migration are still limited.

### 1.1.2 Modeling fish migration

In this paper, we focus on the eDNA dynamics of fish in river environments, which is among the topics where a relatively large amount of knowledge has been accumulated to date. A source of uncertainty in modeling eDNA dynamics is the underlying population dynamics, which are still largely unknown phenomena. Environmental changes such as discharge and water temperature variations have been suggested to affect the migration of major migratory fish species such as salmon and trout [27,28] and some sport fish species [29]. The timing of fish migration has been suggested to be largely scattered across sites and individuals [30] and among years [31,32]. In addition to environmental cues, social interactions among individuals have been suggested to shape fish migration [33]. Empirical catch survey data also introduce uncertainties because catching all migrating fish is impossible [34,35].

Leveraging the versatility of mathematical sciences, the modeling of fish migration has advanced rapidly in recent years; there have been particularly remarkable advances in stochastic models capable of efficiently incorporating diverse uncertainties. A discrete-time stochastic model for identifying evolutionarily stable migration strategies has been formulated and mathematically investigated in detail [36]. A stochastic compartment model of predator–prey interactions has been constructed with a focus on fisheries in the California Current Ecosystem [37]. Interactions between flow and prey aggregation have been described using a stochastic Lagrangian movement model [38]. Fish population dynamics under stochastic ecoepidemiological effects have been analyzed with a focus on their stationarity [39]. A random walk model for fish migration around offshore wind farms has been examined against a real case study [40]. The microscopic schooling behavior of animals, including fish, has been understood to involve a supervised minimization mechanism in random environments [41] and intermittent exploration from visual cues [42]. An advection-diffusion-reaction equation model for migration as well as the stocking and harvesting of multiple species, which potentially applies to fish species, has been developed along with a convergent numerical method [43]. The influences of climate and hydropower changes on the migration of anadromous fish have been investigated through a probabilistic transition model [44]. A stochastic state space model has been employed to clarify changes in cyprinid population dynamics in a freshwater lake [45].

Recently, substantial theoretical results concerning fish migration have been obtained for the upstream migration of *Plecoglossus altivelis altivelis* (Ayu) from seas to rivers. Yoshioka [46] developed a stochastic differential equation (SDE) model for intraday fish counts at a fixed observation point, with which the intermittency of the observed fish count data is explained. Subsequently, the optimality principle behind the SDE model was derived through a Schrödinger bridge formulation [47], and a unified SDE model that can simulate both intraday and daily fish count data was proposed [48]. This series of studies suggested that SDE models with biological backgrounds can capture the upstream fish migration phenomena of *P. altivelis*. More recently, Yoshioka and Louriki [49] extended the SDE model so that the start and end dates of migration can be randomized and presented a simple ODE-based model for studying temporal (but not spatiotemporal) eDNA dynamics for the upstream migration of fish. The model can be made more realistic if the spatiotemporal dynamics of eDNA dynamics are accounted for.

Based on the progress in modeling fish migration phenomena, it would be possible to link this to eDNA modeling via stochastic models; however, such approaches are still limited, and providing a unique case study would fill the gap and advance this research area, motivating the study detailed below.

### 1.2 Aim and contribution

The aims of this study are to formulate and analyze a first-step mathematical model for spatiotemporal eDNA dynamics during seasonal fish migration in river environments. The first contribution of this study is the development of a stochastic partial differential equation (SPDE) with a delayed source for eDNA dynamics. The second contribution is the mathematical analysis and unconditionally stable numerical discretization of this model. The third contribution is its application. Each contribution is explained below.

### 1.2.1 Model formulation: SPDE with a multiplicative noise

We assume a problem setting where one is interested in the upstream migration of a fish along a river, which is considered a 1-D domain (**Figure 1**). As migrants swim upstream against river flow, eDNA particles are shed from their bodies and transported downstream. Fine-scale observation studies have suggested that eDNA concentrations in river environments highly fluctuate [50,51]. Indeed, the shedding of eDNA has conventionally been assumed to scale with fish abundance, but is uncertain [52,53]. Moreover, eDNA may experience biological decay, deposition, and resuspension.

As fully describing these complex and uncertain transport phenomena would face extraordinary difficulties, we consider a simplified situation where the advective transport of migrants and eDNA and decay are explicitly modeled, while the other factors are represented as multiplicative noise proportional to square root of the eDNA concentration:

$$\underbrace{\mathrm{d}Y_t(x)}_{\text{Increment}} = \left( \underbrace{u\frac{\partial Y_t(x)}{\partial x}}_{\text{Transport toward downstream}} + \underbrace{G_t(x)}_{\text{Source}} - \underbrace{R_t(x)}_{\text{Decay}} \right)\mathrm{d}t + (\text{multiplicative noise}), \tag{1}$$

where $t$ is time, $x$ is the 1-D spatial coordinate, $Y_t(x)$ is the eDNA concentration at time $t$ and location $x$, and $u > 0$ is the flow speed where the river flow is in the negative $x$ direction.

In (1), the downstream transport is due to advection along the river. The source does not depend on the eDNA concentration but on the background fish migration, whose temporal dynamics are assumed to be given through an SDE. The source $G_t(x)$ at time $t$ and location $x$ is modeled as a function of $Z_{t-x/v}$, where $Z_t$ represents the unit-time fish count at time $t$ and location $x = 0$, $v$ is the bulk ground speed of migrants (not the instantaneous swimming speed but rather the average migration speed during the migration process), and "$t - x/v$" represents the delayed information such that the shedding at time $t$ and location $x$ is due to migrants that passed the downstream end $x/v$ earlier. The decay $R_t(x)$ is assumed to be proportional to the concentration $Y_t(x)$ motivated by the linear models [e.g., 15,16], but the proportionality coefficient is assumed to be inclusive and is estimated by fitting the model against data.

The multiplicative noise term in (1) represents the factors that are not explicitly represented in the model. We need to design this term so that the eDNA concentration remains nonnegative. Another design criterion of this term is tractability, in such a way that the models are well-defined and can be obtained analytically when necessary. In this view, the SDE models [46-48] offer an important insight that the following square-root multiplicative noise potentially leads to an SPDE model whose Laplace functional is found in a closed form: $\sigma\sqrt{Y_t(x)}W(\mathrm{d}t,\mathrm{d}x)$ with $\sigma > 0$ being noise intensity and $W$ representing space-time white noise (e.g., Definition 1.3.1 in Dalang and Sanz-Solé [54]). This choice is also based on an analogy with the fluctuating nature of biological processes, such as upstream fish migration, as a continuous-time branching process [46]. We therefore use this formulation.

An important topic where SPDEs related to (1) arise is the yield curve of the option price, with $x$ remaining until maturity, often called Heath–Jarrow–Morton models [55,56]. In fact, as discussed later, the proposed model is closely related to SP(D)E in finance. In biology, SPDEs with square-root multiplicative noise have been employed for modeling biological and ecological phenomena such as phytoplankton aggregation [57,58], which also motivated the use of our noise term. Finally, in the context of fish migration, SPDEs for the random occurrence of fish have been proposed by Kanamori et al. [59]. An inverse problem of SPDEs motivated by the source reconstruction of eDNA concentrations has been discussed in Okasaki et al. [60]. These SPDEs have different forms from ours but are considered related models that focus on fish migration.

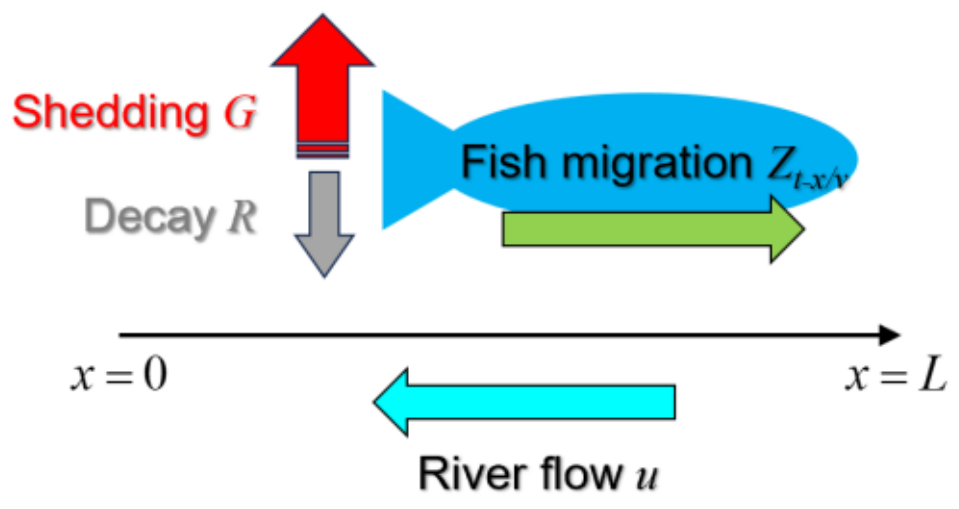

**Figure 1.** Conceptual diagram for the modeling in this paper.

### 1.2.2 Mathematical analysis and numerical discretization

The square-root multiplicative noise in (1) combined with the specific form of the SDE model [48] allows for the closed-form derivation of the Laplace functional of the eDNA concentration. In this case, the proposed SPDE model is an infinite-dimensional (measure-valued) affine process [61]; our model additionally possesses distributed delay in the source term. Loosely speaking, an (nonnegative) affine diffusion process has a linear drift term combined with square-root multiplicative noise. We exploit knowledge about affine diffusion processes with delay [62,63] to estimate and verify the Laplace functional of the eDNA concentration. The affine nature motivates us to see the eDNA concentration as a measure-valued process whose Laplace functional is found semianalytically (e.g., Chapter 4 in Rudnicki and Wieczorek [64]).

We also consider a semidiscrete version of the SPDE model (1) where the spatial domain is discretized by grid points. This semidiscrete model is a finite-dimensional affine process whose solution exists uniquely in a pathwise sense, and its limit is expected to be a solution to the SPDE model. We show that the Laplace functional of the semidiscrete model is well-defined and that its limit is the Laplace functional of the SPDE model. To address the existence issue, we apply a tightness argument for measure-valued processes (Theorem II.4.1 of Perkins [65]) to the semidiscrete model, showing that its continuum limit is the SPDE model in a weak sense. This observation, combined with the convergence results between the Laplace functionals (Theorem 1.19 in Li [66]), yields that the limit is unique in law.

From an engineering standpoint, time discretization of the semidiscrete model needs care because the coefficient $\sqrt{Y_t(x)}$ in the noise term may become computationally indefinite, particularly when the noise intensity is high. We exploit the recently developed numerical scheme for square-root diffusion processes [67] and apply it to the semidiscrete model, thereby unconditionally guaranteeing the nonnegativity of numerical solutions.

Affine processes with delay have been studied both theoretically and computationally in the context of stochastic Volterra processes, where nonexponential kernel functions represent memory characteristics and pathwise roughness [68,69]. Their application studies, particularly those targeting finance and insurance, have also been addressed [70,71]. The proposed SPDE model has different forms from these models, although both have closed-form Laplace functionals. From a computational standpoint, nonnegativity-preserving numerical solutions of S(P)DEs with square-root diffusion coefficients have been addressed by considering exact sampling [72] and adaptive time-stepping [73]. The scheme of Abi Jaber [67] is simpler to implement and employs a one-step time discretization, as in the classical Euler–Maruyama method, without any adaptive stepping or complex sampling. Computational studies about measure-valued solutions are still rare.

#### 1.2.3 Application case study

Our application study aims to investigate the model (1) on the basis of eDNA concentration data for the upstream migration of *P. altivelis* in a river environment in Japan. We identify the coefficients in the model based on a least-squares fit between modeled and observed eDNA concentrations. We computationally investigate the behavior of the proposed model and perform a sensitivity analysis to obtain deeper insights into the upstream migration of *P. altivelis*.

### 1.3 Structure of this paper

The rest of this paper is structured as explained below. **Section 2** presents our problem setting and model. **Section 3** theoretically investigates the proposed model. **Section 4** presents an application case study. **Section 5** summarizes this study and presents future perspectives. **Appendix** presents proofs and technical results (**Sections A1-A2**) and supporting data (**Section A3**).

## 2. Mathematical model

The SPDE model and its semidiscrete version are presented. We will work on a complete filtered probability space $\left(\Omega,\mathcal{F},\left(\mathcal{F}_t\right)_{t\geq 0},\mathbb{P}\right)$ as usual in stochastic calculus, and $\mathbb{E}$ represents the expectation. Subscripts and arguments of variables are omitted when there is no confusion.

## 2.1 SPDE model

We fully formulate our SPDE model which has been half-explained in **Section 1.2.1**. We consider the spatiotemporal dynamics of eDNA along a river modeled as a 1-D domain $D=(0,L)$ with $L>0$ (**Figure 1**). We write $\bar{D}=[0,L]$. The direction of river flow is from $x=L$ (upstream-end) to $x=0$ (downstream-end). The eDNA concentration of a target fish species at time $t$ and location $x$ is denoted by $Y_t(x)$. The speed of river flow is a constant $u>0$ for modeling simplicity. We assume that the fish migrate upstream along $D$ with a ground speed $v>0$.

The unit-time fish count is $Z_t$ ($t\in\mathbb{R}$) at $x=0$, where we assume $Z_t=0$ ($t<0$), meaning that the fish migration in $D$ occurs for $t\geq 0$. The deterministic shedding of eDNA from migrants at time $t$ and location $x$ is assumed to be $G_t(x)=g_t(x)Z_{t-x/v}$, where $g_t(x)>0$ is the shedding rate and the factor $Z_{t-x/v}$ means that the shedding at $(t,x)$ is due to the migrant that passed $x=0$ by $x/v$ ago. The deterministic eDNA decay at $(t,x)$ is $R_t(x)=r_t(x)Y_t(x)$, where $r_t(x)>0$ is a decay rate. The remaining uncertainties in the eDNA dynamics are modeled by the square-root multiplicative noise $\sigma\sqrt{Y_t(x)}W(\mathrm{d}t,\mathrm{d}x)$, where $\sigma_t(x)>0$ is the noise intensity and $W$ is space-time white noise with the following formal covariance structure (e.g., Definition 1.3.1 in Dalang and Sanz-Solé [54]):

$$\mathbb{E}\left[W(\mathrm{d}t,\mathrm{d}x)W(\mathrm{d}s,\mathrm{d}y)\right]=\delta(x-y)\delta(t-s),\quad s,t>0,\quad x,y\in D, \tag{2}$$

where $\delta(\cdot)$ is Dirac's delta.

Our SPDE model is formulated as follows:

$$\mathrm{d}Y_t(x)=\left(u\frac{\partial Y_t(x)}{\partial x}+g_t(x)Z_{t-x/v}-r_t(x)Y_t(x)\right)\mathrm{d}t+\sigma_t(x)\sqrt{Y_t(x)}W(\mathrm{d}t,\mathrm{d}x),\quad t>0,\quad x\in D \tag{3}$$

subject to the initial condition $Y_0(x)=0$ for $x\in D$ (there are no target fish in $D$ at time 0) and boundary condition $Y_t(L)=0$ for $t>0$ (migrants cannot pass through the upstream-end $x=L$). No boundary condition is specified at $x=0$ as there is a free outflow of eDNA from this boundary. In the present model, the source is assumed to vanish at $Y_t(L)=0$, which is an assumption. In reality, however, they would establish a territory somewhere before reaching the upstream end. Since this study focuses exclusively on the migratory phenomenon of upstream movement, this aspect has been simplified.

Few remarks about the SPDE model (3) are in order.

***Remark 1.*** Theoretically, the domain $D=(0,L)$ can be replaced by an unbonded domain such as $[0,+\infty)$, although this is less realistic. In this case, we need some decay estimates for solutions for $x=+\infty$.

***Remark 2.*** Noise whiteness is assumed to simplify the model, and a spatial correlation structure, which possibly depends on both fish migration dynamics and river hydraulics, may be considered. This topic is beyond the scope of this paper and will be addressed in the future.

***Remark 3.*** In applications, the flow speed of river and ground speed of fish would depend on space and time. In this case, however, the analysis carried out in this paper becomes more complicated.

### 2.2 Semidiscrete model

We also consider a semidiscrete version of the SPDE model (3). The domain is discretized into equidistant grid points, the solution and coefficients on them are evaluated, and the spatial partial differential is replaced with a first-order difference. The total number of grid points is $N \in \mathbb{N}$, and we write $I = \{1, 2, 3, ..., N\}$. The $i$ th grid point is $x_i = (i - 1/2)h$ with $h = L/N$, and the quantities evaluated on $x_i$ are expressed with the superscript $i$, e.g., $Y_t^{(i)}$. We set $\tau_i = x_i / v > 0$ ($i \in I$).

The semidiscrete model is formulated as follows, where we formally set $Y_t^{(N+1)} = 0$:

$$\mathrm{d}Y_t^{(i)} = \left( u \frac{Y_t^{(i+1)} \mathbb{I}(i < N) - Y_t^{(i)}}{h} + g_t^{(i)} Z_{t-\tau_i} - r_t^{(i)} Y_t^{(i)} \right) \mathrm{d}t + \sigma_t^{(i)} \sqrt{Y_t^{(i)}} \frac{\mathrm{d}W_t^{(i)}}{\sqrt{h}}, \quad t > 0, \quad i \in I \tag{4}$$

subject to the initial condition $Y_0^{(i)} = 0$ for $i \in I$; for the discretization of the noise term, see El Saadi and Bah [58]. Here, $\mathbb{I}(S)$ is the indicator function for the set $S$ ($\mathbb{I}(S) = 1$ if $S$ is true and $\mathbb{I}(S) = 0$ otherwise), and $\left(W_t^{(i)}\right)_{t \geq 0}$ ($i \in I$) are independent 1-D standard Brownian motions. The system (4), provided that $(Z_t)_{t \geq 0}$ is an affine process, is also an affine process (with delay). This fact plays a role in the theoretical and numerical analysis in the next section.

The semidiscrete (4) model is considered a finite volume spatial discretization of the SPDE model (3), where the spatial partial derivative is replaced by upwind finite difference (e.g., Grün et al. [74]). Computationally, it is possible to employ a higher-order discretization method for this term; however, such a method is not monotone and cannot maintain the nonnegativity of solutions to (4) even without noise or results in a highly nonlinear drift term that is not tractable [75]. The upwind discretization is therefore essential in this study.

## 3. Mathematical analysis and discretization

We mathematically investigate the SPDE model and its semidiscrete version. We also present a fully discretized version to be used in numerical computation.

In the rest of this paper, we assume that the coefficients $g_t(x), r_t(x), \sigma_t(x)$ are positive, bounded, and continuously differentiable for all $t \geq 0$ and $x \in \bar{D}$, and we fix $T > 0$. The collection of nonnegative measures on $D$ is denoted by $\mathcal{M}(D)$. In what follows, a measure on $D$ is regarded as a measure on $[0, L)$ whenever convenient, which does not change the measure itself but allows the use of the compact sets $[0, L-1/l]$ ($l > L^{-1}$) of $[0, L)$. The Borel sigma-algebra of $D$ is denoted by $\mathcal{B}(D)$. We call a process $\mu : [0,T] \times \mathcal{B}(D) \times \Omega \to [0,\infty)$ a measure-valued process if, for any fixed time $t \in [0,T]$ and sample path $\omega \in \Omega$, $\mu_t(\cdot, \omega) \in \mathcal{M}(D)$ and is cadlag. We omit the dependence on $\omega$ in the sequel. The collection of measure-valued processes endowed with the Skorohod topology is denoted by $\mathbb{D}([0,T], \mathcal{M}(D))$. The collection of functions $\phi : \bar{D} \to \mathbb{R}$ that are continuously differentiable arbitrary times on $\bar{D}$ is denoted by $C^{\infty}(\bar{D})$. For any $\phi : \bar{D} \to \mathbb{R}$, we set the norm $\|\phi\|_{\infty} = \sup_{x \in \bar{D}} |\phi(x)|$. With an abuse of notation, for any $\varphi : [0,+\infty) \times \bar{D} \to \mathbb{R}$, we set $\|\varphi\|_{\infty} = \sup_{0 \le t \le T, x \in \bar{D}} |\varphi_t(x)|$.

### 3.1 Unit-time fish count

In the rest of this paper, we assume that the unit-time fish count $Z$ follows the SDE below, which applies at least to *P. altivelis* [48]:

$$\mathrm{d}Z_t = \left( a_t - \frac{b}{\tau - t} Z_t \right) \mathrm{d}t + c_t \sqrt{\frac{b}{\tau - t} Z_t} \, \mathrm{d}B_t , \quad 0 < t < \tau \tag{5}$$

with initial and terminal conditions $Z_0 = Z_\tau = 0$. Here, $a_t, c_t$ are nonnegative, bounded, and continuously differentiable functions of $t \in [0, \tau]$, $b > 0$ is a constant, $\tau > 0$ is the terminal time of migration, and $(B_t)_{0 \le t \le \tau}$ is a 1-D standard Brownian motion independent of $W$ and $\left(W_t^{(i)}\right)_{t \ge 0}$ ($i \in I$). We extend $(Z_t)_{0 \le t \le \tau}$ to $(Z_t)_{t \in \mathbb{R}}$ by setting $Z_t = 0$ for $t < 0$ and $t > \tau$. The SDE (5) is of an affine type and admits a unique, nonnegative, square-integrable, and pathwise solution in $[0, \tau]$ such that $Z_0 = Z_\tau = 0$ with probability 1 (see **Proof of Proposition 1** in **Appendix**). Note that this SDE is related to a Brownian bridge because its initial and terminal conditions are pinned.

***Remark 4.*** The start and end of migration may be randomized in applications [49] but are fixed in this study to maintain the analytical tractability of the proposed model.

### 3.2 Well-posedness

#### 3.2.1 Semidiscrete model

The semidiscrete model (4) is an SDE system with non-Lipschitz diffusion coefficients. We can effectively consider this model as a cascading system of SDEs that can be solved from $i = N$ to $i = 1$ because of the upwind discretization. Its well-posedness is obtained below.

***Proposition 1***

*The SDE system (4) admits a unique pathwise solution* $\left(Y_t^{(i)}\right)_{t\geq 0}$ *(* $i \in I$ *) whose components are nonnegative and continuous with probability 1.*

### 3.2.2 SPDE model

Since the SPDE model (3) may not have smooth solutions, we define its weak version. Consider any test function $\phi \in C^{\infty}\left(\bar{D}\right)$. Multiplying $\phi$ by (3), integrating over $D$, and integrating by parts yields

$$\begin{aligned}\int_0^L \phi(x)\mathrm{d}Y_t(x)\mathrm{d}x = &\left(-u\phi(0)Y_t(0) - \int_0^L u\frac{\mathrm{d}\phi(x)}{\mathrm{d}x}Y_t(x)\mathrm{d}x + \int_0^L \phi(x)\left(g_t(x)Z_{t-x/v} - r_t(x)Y_t(x)\right)\mathrm{d}x\right)\mathrm{d}t \\ &+\int_0^L \phi(x)\sigma_t(x)\sqrt{Y_t(x)}W(\mathrm{d}t,\mathrm{d}x).\end{aligned} \tag{6}$$

Here, the non-inclusion of the boundary term at $x = L$ in equation (6) is because of the assumption $Y_t(L) = 0$. The term $-u\phi(0)Y_t(0)$ corresponds to the free outflow of eDNA. This, along with the initial condition $Y_0(x) = 0$, shows that the process $\left(M_t(\phi)\right)_{t\geq 0}$ depending on $\phi$ is a zero-mean martingale if the right-hand side of the equation below exists:

$$\begin{aligned}M_t(\phi) = &\int_0^L \phi(x)Y_t(x)\mathrm{d}x + \int_0^t u\phi(0)Y_s(0)\mathrm{d}s \\ &-\int_0^t\left(-\int_0^L u\frac{\mathrm{d}\phi(x)}{\mathrm{d}x}Y_s(x)\mathrm{d}x + \int_0^L \phi(x)\left(g_s(x)Z_{s-x/v} - r_s(x)Y_s(x)\right)\mathrm{d}x\right)\mathrm{d}s.\end{aligned} \tag{7}$$

The quadratic variation $\langle M_t(\phi)\rangle$ of $M_t(\phi)$ should be given by

$$\langle M_t(\phi)\rangle = \int_0^t\int_0^L \left(\phi(x)\right)^2\left(\sigma_s(x)\right)^2 Y_s(x)\mathrm{d}x\mathrm{d}s. \tag{8}$$

We suspect that solutions to the SPDE model are measure-valued processes because of the lack of smoothing terms such as the Laplacian in stochastic heat equations (Chapter 3 in Dalang and Sanz-Solé [54]) and 1-D super-Brownian motions (Chapter 4 in Rudnicki and Wieczorek [64]). In this view, we introduce a measure version $\mu_t(\mathrm{d}x) = Y_t(x)\mathrm{d}x$ so that $M_t(\phi)$ is rewritten as

$$\begin{aligned}M_t(\phi) = &\int_0^L \phi(x)\mu_t(\mathrm{d}x) + \int_0^t u\phi(0)Y_s(0)\mathrm{d}s \\ &-\int_0^t\left(-\int_0^L u\frac{\mathrm{d}\phi(x)}{\mathrm{d}x}\mu_s(\mathrm{d}x) + \int_0^L \phi(x)g_s(x)Z_{s-x/v}\mathrm{d}x - \int_0^L \phi(x)r_s(x)\mu_s(\mathrm{d}x)\right)\mathrm{d}s\end{aligned} \tag{9}$$

with the quadratic variation

$$\langle M_t(\phi)\rangle = \int_0^t\int_0^L \left(\phi(x)\right)^2\left(\sigma_s(x)\right)^2 \mu_s(\mathrm{d}x)\mathrm{d}s. \tag{10}$$

Hereafter, $\mu \in \mathbb{D}\left(\left[0,T\right],\mathcal{M}(D)\right)$ is called a weak solution (of martingale problem) to the SPDE model (3) if $M_t(\phi)$ is a zero-mean martingale satisfying (10) for any $\phi \in C^{\infty}\left(\bar{D}\right)$. Note that weak solutions deal with the white noise only implicitly.

***Remark 5.*** Any solution ($Y$ or $\mu$) is determined from upstream to downstream, and hence the free-outflow condition is determined from the information in $D$. This solution structure is inherited in the semidiscrete model and numerical scheme.

### 3.3 Laplace functional

#### 3.3.1 Semidiscrete model

The Laplace functional of the semidiscrete model (4) is set as follows: for $\lambda^{(i)} \geq 0$ ($i \in I$),

$$C_N(t,T) = \mathbb{E}\left[\exp\left(-h\sum_{i=1}^{N}\lambda^{(i)}Y_T^{(i)}\right)\middle|\mathcal{F}_t\right],\quad 0 \leq t \leq T . \tag{11}$$

The following proposition gives a closed-form formula for $C_N(t,T)$.

***Proposition 2***

*The Laplace functional (11) is expressed as*

$$C_N(t,T) = \exp\left(\alpha_{N,t} + \beta_{N,t}Z_t + h\sum_{i=1}^{N}\gamma_{N,t}^{(i)}Y_t^{(i)} + h\sum_{i=1}^{N}\int_{t-\tau_i}^{t}\omega_{N,s,t}^{(i)}Z_s\mathrm{d}s\right),\quad 0 \leq t \leq T , \tag{12}$$

*where the coefficients $\alpha_N, \beta_N, \gamma_N, \omega_N$ in (12) are continuous in time and are uniquely determined from the following system: for $t < T$,*

$$\frac{\mathrm{d}\alpha_{N,t}}{\mathrm{d}t} + \beta_{N,t}a_t = 0 , \tag{13}$$

$$\frac{\mathrm{d}\beta_{N,t}}{\mathrm{d}t} - \frac{b}{\tau - t}\beta_{N,t} + \frac{1}{2}\frac{b(c_t)^2}{\tau - t}\left(\beta_{N,t}\right)^2 + h\sum_{i=1}^{N}\omega_{N,t,t}^{(i)} = 0,\quad t < \tau , \tag{14}$$

$$\frac{\mathrm{d}\gamma_{N,t}^{(i)}}{\mathrm{d}t} - u\frac{\gamma_{N,t}^{(i)} - \gamma_{N,t}^{(i-1)}}{h} - r_t^{(i)}\gamma_{N,t}^{(i)} + \frac{1}{2}\left(\gamma_{N,t}^{(i)}\right)^2\left(\sigma_t^{(i)}\right)^2 = 0,\quad i > 1 , \tag{15}$$

$$\frac{\mathrm{d}\gamma_{N,t}^{(1)}}{\mathrm{d}t} - u\frac{\gamma_{N,t}^{(1)}}{h} - r_t^{(1)}\gamma_{N,t}^{(1)} + \frac{1}{2}\left(\gamma_{N,t}^{(1)}\right)^2\left(\sigma_t^{(1)}\right)^2 = 0 , \tag{16}$$

*and*

$$\omega_{N,t,t}^{(i)} = \begin{cases} 0 & \left(T - \tau_i \leq t < T\right) \\ \gamma_{N,t+\tau_i}^{(i)}g_{t+\tau_i}^{(i)} & \left(t < T - \tau_i\right) \end{cases},\quad i \in I \tag{17}$$

*along with the terminal condition*

$$\gamma_{N,T}^{(i)} = -\lambda^{(i)} \leq 0 \text{ for } i \in I,\ \alpha_{N,T} = 0,\ \beta_{N,t} = 0\ (\min\{\tau,T\} \leq t \leq T),\ \omega_{N,T,T}^{(i)} = 0 \text{ and } i \in I . \tag{18}$$

#### 3.3.2 SPDE model

As for the semidiscrete case, the Laplace functional of the model (3) is set as follows: for nonnegative $\lambda \in C^{\infty}\left(\bar{D}\right)$,

$$C(t,T)=\mathbb{E}\left[\exp\left(-\int_0^L \lambda(x)\mu_T(\mathrm{d}x)\right)\middle|\mathcal{F}_t\right],\ \ 0\le t\le T\ . \tag{19}$$

We have the continuous version of **Proposition 2**.

***Proposition 3***

*The Laplace functional in (19) can be expressed as*

$$C(t,T)=\exp\left(\alpha_t+\beta_t Z_t+\int_0^L \gamma_t(x)\mu_t(\mathrm{d}x)+\int_0^L\left(\int_{t-x/v}^t \omega_{s,t}(x)Z_s\mathrm{d}s\right)\mathrm{d}x\right),\ \ 0\le t\le T\ , \tag{20}$$

*where the coefficients* $\alpha,\beta,\gamma,\omega$ *in (20) are determined from the following system: for* $t<T$ *,*

$$\frac{\mathrm{d}\alpha_t}{\mathrm{d}t}+\beta_t a_t=0, \tag{21}$$

$$\frac{\mathrm{d}\beta_t}{\mathrm{d}t}-\frac{b}{\tau-t}\beta_t+\frac{1}{2}\frac{b(c_t)^2}{\tau-t}(\beta_t)^2+\int_0^L \omega_{t,t}(x)\mathrm{d}x=0,\ \ t<\tau, \tag{22}$$

$$\frac{\partial\gamma_t(x)}{\partial t}-u\frac{\partial\gamma_t(x)}{\partial x}-r_t(x)\gamma_t(x)+\frac{1}{2}(\gamma_t(x))^2(\sigma_t(x))^2=0,\ \ x\in D, \tag{23}$$

*and*

$$\omega_{t,t}(x)=\begin{cases}0 & (T-x/v\le t<T)\\ \gamma_{t+x/v}(x)g_{t+x/v}(x) & (t<T-x/v)\end{cases}, \tag{24}$$

*along with the boundary condition*

$$\gamma_t(0)=0\ \ \textit{for}\ \ t<T \tag{25}$$

*and the terminal condition*

$$\gamma_T(x)=-\lambda(x)\le 0\ \ \textit{for}\ \ x\in D,\ \ \alpha_T=0,\ \ \beta_t=0\ \ (\min\{\tau,T\}\le t\le T\ ),\ \ \omega_{T,T}(x)=0\ \ \textit{for}\ \ x\in D. \tag{26}$$

### 3.4 Convergence results

#### 3.4.1 Convergence of Laplace functional

The law of $\mu$ is uniquely determined by **Proposition 2**. This follows if $C_N(t,T)\underset{N\to+\infty}{\to}C(t,T)$ pointwise by Theorem 1.19 in Li [66] for the one-to-one relationship between Laplace functionals and probability laws. To proceed, we set

$$\mu_{N,t}(\mathrm{d}x)=h\sum_{i=1}^N Y_t^{(i)}\delta(x-x_i)\ . \tag{27}$$

We have the following proposition.

***Proposition 4***

*From (12)-(20),* $C_N(t,T)\underset{N\to+\infty}{\to}C(t,T)$ *pointwise at each* $t\in[0,T]$*; hence, the Laplace functional of the SPDE model (3) must have the form of (20).*

### 3.4.2 Existence and uniqueness

Regarding the existence of a weak solution to the SPDE model, for any $\phi \in C^{\infty}\left(\bar{D}\right)$, consider

$$\tilde{Y}_{N,t}\left(\phi\right) = \int_0^L \phi\left(x\right)\mu_{N,t}\left(\mathrm{d}x\right) = h\sum_{i=1}^{N} Y_t^{(i)}\phi\left(x_i\right),\ \ t \geq 0 . \tag{28}$$

Applying Itô's formula to $\tilde{Y}_{N,t}\left(\phi\right)$ yields the semidiscrete version of (9):

$$M_{N,t}\left(\phi\right) = \tilde{Y}_{N,t}\left(\phi\right) - \int_0^t \left( -uh\sum_{i=1}^{N} Y_s^{(i)} \frac{\phi\left(x_i\right) - \mathbb{I}\left(i>1\right)\phi\left(x_{i-1}\right)}{h} + h\sum_{i=1}^{N} g_s^{(i)} Z_{s-x_i/v}\phi\left(x_i\right) - h\sum_{i=1}^{N} r_s^{(i)} Y_s^{(i)}\phi\left(x_i\right) \right)\mathrm{ds}. \tag{29}$$

Here, we used

$$\sum_{i=1}^{N}\phi\left(x_i\right)\frac{Y_s^{(i+1)}\mathbb{I}\left(i<N\right) - Y_s^{(i)}}{h} = -\sum_{i=1}^{N} Y_s^{(i)} \frac{\phi\left(x_i\right) - \mathbb{I}\left(i>1\right)\phi\left(x_{i-1}\right)}{h}, \tag{30}$$

where we formally set $\phi\left(x_0\right) = 0$. The quadratic variation $\left\langle M_{N,t}\left(\phi\right)\right\rangle$ of $M_{N,t}\left(\phi\right)$ is

$$\left\langle M_{N,t}\left(\phi\right)\right\rangle = \int_0^t h\sum_{i=1}^{N} Y_s^{(i)}\left(\phi\left(x_i\right)\right)^2\left(\sigma_s^i\right)^2 \mathrm{ds} . \tag{31}$$

First, we show that the process

$$\mathbb{Y}_{N,t}\left(\phi\right) = \tilde{Y}_{N,t}\left(\phi\right) + \phi\left(0\right)\mathbb{F}_{N,t} \ \text{ with } \ \mathbb{F}_{N,t} = u\int_0^t Y_s^{(1)}\mathrm{ds} \tag{32}$$

satisfies the Aldous criterion. Here, $\mathbb{F}_{N,t}$ is the cumulative amount of eDNA that has flowed out of the domain through $x = 0$ until time $t$.

***Proposition 5***

*For any $\phi \in C^{\infty}\left(\bar{D}\right)$, the sequence of processes $\left(\mathbb{Y}_{N,t}\left(\phi\right)\right)_{0\leq t\leq T}$ ($N = 1,2,3,...$) satisfies the following Aldous criterion; for all $\varepsilon > 0$ and $\zeta > 0$, there are some $\eta > 0$ and $N_0 \in \mathbb{N}$ such that*

$$\sup_{N\geq N_0}\ \sup_{0\leq S\leq S'\leq \min\{S+\eta,T\}} \mathbb{P}\left(\left|\mathbb{Y}_{N,S}\left(\phi\right) - \mathbb{Y}_{N,S'}\left(\phi\right)\right| > \varepsilon\right) \leq \zeta , \tag{33}$$

*where $S, S'$ are stopping times such that $S \leq S'$.*

Second, we show an additional result about $\left(\mu_{N,t}\right)_{0\leq t\leq T}$.

***Proposition 6***

*For each $l > L^{-1}$, set $D_l = \left[L - 1/l, L\right)$. For any $\eta > 0$, there exists some $l > 0$ such that*

$$\sup_{N\in\mathbb{N}} \mathbb{P}\left( \sup_{0\leq t\leq T}\mu_{N,t}\left(D_l\right) > \eta \right) < \eta . \tag{34}$$

By **Propositions 5-6**, we obtain the existence result of weak solutions to the SPDE model (3). Also, by the convergence result of the Laplace functional, we obtain the convergence from the semidiscrete model to the SPDE model in law.

***Proposition 7***

*The SPDE model (3) admits a weak solution. Moreover, weak solutions are unique in law.*

Obtaining the well-posedness in a stronger sense (e.g., pathwise sense) is considered difficult for the proposed model because of using the notion of weak solutions based on martingale problems.

### 3.5 Numerical discretization

We present a fully discrete version of the SPDE model (3), which is the time discretization of the semidiscrete model (4). We use an operator-splitting method, which separately handles deterministic and stochastic terms. We set a time grid with the time instances $t_j = j\rho$ $(j = 0,1,2,...)$ with a time increment $\rho > 0$. The discretized quantity at time $t_j$ is represented by the superscript $j$. The initial condition is $Y^{(i,0)} = 0$ ($i \in I$). We assume that the process $Z_t$ has already been discretized in time by using a numerical scheme that guarantees nonnegativity; in this paper, we assume the discretization of Yoshioka (2025)[46]. Our numerical method has the following two steps.

#### 3.5.1 Deterministic terms

In the first step, we apply an implicit Euler discretization to the deterministic terms as follows:

$$\begin{gathered}
\mathrm{d}Y_t^{(i)} = \left( u \frac{Y_t^{(i+1)} \mathbb{I}(i < N) - Y_t^{(i)}}{h} + g_t^{(i)} Z_{t-\tau_i} - r_t^{(i)} Y_t^{(i)} \right) \mathrm{d}t \\
\Downarrow \\
\hat{Y}^{(i,j+1)} - Y^{(i,j)} = \left( u \frac{\hat{Y}^{(i+1,j+1)} \mathbb{I}(i < N) - \hat{Y}^{(i,j+1)}}{h} + g^{(i,j)} Z_{t_j - \tau_i} - r^{(i,j)} \hat{Y}^{(i,j+1)} \right) \rho
\end{gathered} \tag{35}$$

and hence (we set $Y^{(N+1,\cdot)} = \hat{Y}^{(N+1,\cdot)} = 0$)

$$\hat{Y}^{(i,j+1)} = \left( 1 + \left( \frac{u}{h} + r^{(i,j)} \right) \rho \right)^{-1} \left( Y^{(i,j)} + \frac{u \mathbb{I}(i < N)}{h} \hat{Y}^{(i+1,j+1)} \rho + g^{(i,j)} Z_{t_j - \tau_i} \rho \right), \tag{36}$$

which is explicitly solvable from $i = N$ to $i = 1$. By (36), the nonnegativity of $\hat{Y}^{(i,j+1)}$ ($i \in I$) when $Y^{(i,j)} \geq 0$ ($i \in I$) is guaranteed unconditionally irrespective of $\rho > 0$.

#### 3.5.2 Other terms

In the second step, we integrate the following system of decoupled SDEs starting from $\hat{Y}^{(i,j+1)}$:

$$\mathrm{d}Y_t^{(i)} = \frac{\sigma^{(i,j)}}{\sqrt{h}}\sqrt{Y_t^{(i)}}\mathrm{d}W_t^{(i)}, \quad t_j < t < t_{j+1}, \quad i \in I. \tag{37}$$

This is carried out by using the scheme [67], which theoretically generates nonnegative solutions to (37). More specifically, we discretize (37) as follows:

$$Y^{(i,j+1)} = \hat{Y}^{(i,j+1)} + \frac{\sigma^{(i,j)}}{\sqrt{h}}V^{(i,j)} \left(= \rho^{-1}U^{(i,j)}\right), \quad i \in I, \tag{38}$$

where

$$\phi^{(i,j)} = \hat{Y}^{(i,j+1)}\rho \text{ and } \psi^{(i,j)} = \frac{\sigma^{(i,j)}}{\sqrt{h}}\rho \tag{39}$$

and

$$U^{(i,j)} \sim \mathrm{IG}\left(\phi^{(i,j)}, \left(\phi^{(i,j)} / \psi^{(i,j)}\right)^2\right) \text{ and } V^{(i,j)} = \frac{1}{\psi^{(i,j)}}\left(U^{(i,j)} - \phi^{(i,j)}\right). \tag{40}$$

Here, $\mathrm{IG}(\mu, \lambda)$ represents the inverse Gaussian distribution whose probability density function is proportional to $z^{-3/2}e^{-\frac{\lambda(z-\mu)^2}{2\mu^2 z}}$ ($z > 0$), and each $U^{(i,j)}$ is generated independently (Algorithm 2 in Abi Jaber [67]). We set $U^{(i,j)} = 0$ when $\hat{Y}^{(i,j+1)} = 0$.

The theoretical nonnegativity of $Y^{(i,j+1)}$ directly follows from Theorem 1.3 in Abi Jaber [67]. We have the following proposition about numerical solutions.

***Proposition 8***

*All* $Y^{(i,j)}$ *(*$i \in I$, $j = 0,1,2,...$*) are nonnegative.*

## 4. Application study

An application study of the proposed SPDE model is presented.

### 4.1 Study site

The study site is the Hii River, which is a Class A river that flows through the eastern part of Shimane Prefecture and its tributary, the Mitoya River (**Figure 2**). The target fish species is the famous diadromous fish species *P. altivelis* in Japan, which has a one-year life cycle (and hence does not have any age structures) and migrates between rivers and seas every year. For the detailed biology and ecology of *P. altivelis*, see Tsukamoto and Uchida [76]. Our target is the spring upstream migration of juvenile fish that occurs from March to July every year in the Hii River system. Because the fish grow and spawn in this river from July to October and November, investigations of their upstream migration are crucial for assessing their population dynamics in this river.

The sampling sites are Negai Bridge in the mainstream Hii River (called Kisuki: 35°17'35"N, 132°53'51"E) and Kyu-Shita Bridge in the tributary Mitoya River (called Shin-Mitoya: 35°18′06.87″N, 132°53′23.43″E), as shown in **Figure 2**. We collected the eDNA concentration of *P. altivelis* weekly from April to October in 2025. Laboratory analysis of eDNA was performed with species-specific real-time PCR based on the primers listed in Table 1 in Yamanaka and Minamoto [77] (see **Tables A1-A2**) along with Environmental DNA Sampling and Experiment Manual Ver 3.0 (Chapter 5 in The eDNA Society [78]). We employed glass fiber filter paper with an aperture size of 0.6 (μm). We used a DNeasy Blood & Tissue Kit (QIAGEN) for DNA extraction and the THUNDERBIRD Probe qPCR Mix (TOYOBO) with a hydrolysis probe in the laboratory analysis.

The eDNA concentration data obtained are shown in **Figure 3**, where the first peak corresponds to upstream migration, while the second peak corresponds to spawning migration; we focus on the former in this paper. See also **Table A3** for the corresponding numerical data. The observed data suggest that the eDNA concentration is higher in the tributary than in the mainstream, which is considered due to the higher attraction of the former having a higher flow speed, as suggested in the next subsection.

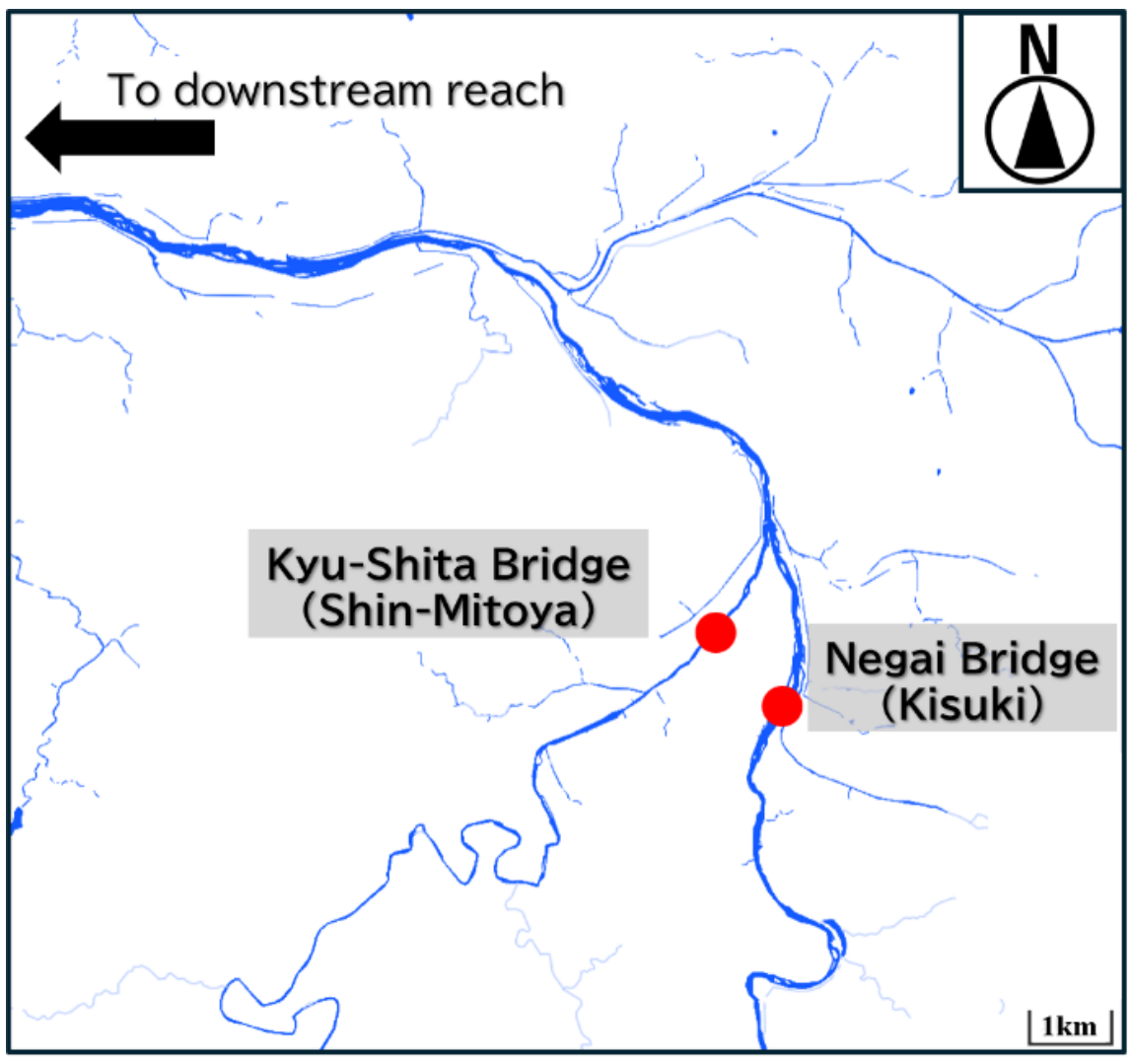

**Figure 2.** Map of the study site.

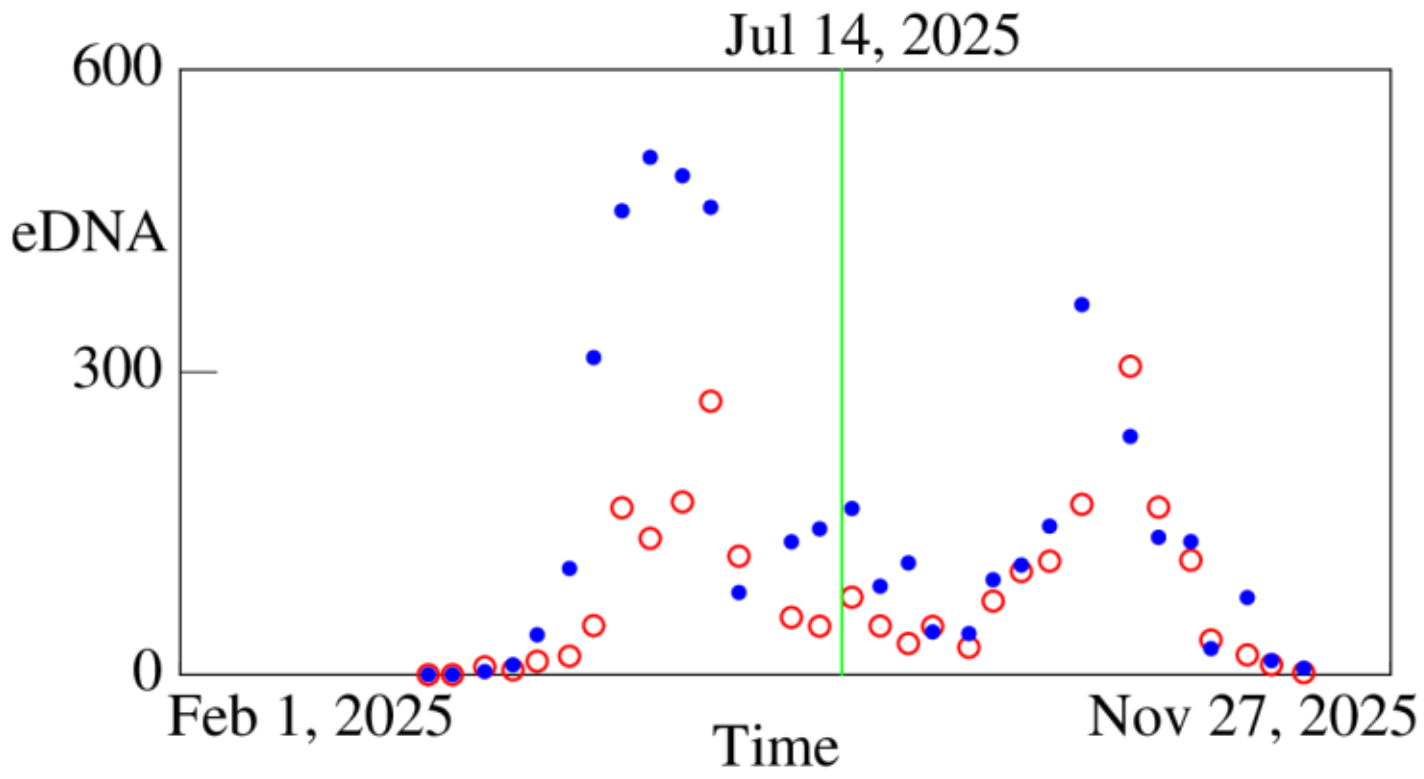

**Figure 3.** Collected eDNA data from February 1 to November 27: Kisuki (red unfilled circles) and Shin-Mitoya (blue filled circles).

### 4.2 Parameter estimation

The objective here is to demonstrate the model's operation, while parameter identification using high-frequency data is reserved for future work. In the rest of this paper, we assume that the density $Y$ of $\mu$ exists. The parameter identification for the process $Z$ follows that in Yoshioka and Louriki[49]; based on their threshold water temperature analysis at Kisuki, they identified the migration duration of $\tau = 127$ (day) in the downstream part from middle March (water temperature in the midstream exceeds approximately 9 (°C)) to middle July (water temperature in the midstream exceeds approximately 23 (°C)). However, time 0, namely, the initiation of the upstream migration in the midstream, is uncertain unless some other assumption is specified, which is therefore optimized here. This study also uses their estimation results. Time 0 is therefore estimated here as well.

The coefficients $a_t$ and $c_t$ in the SDE (5) are set following Yoshioka [48] who proposed to first model cumulants and then consistently set coefficients:

$$a_t = \frac{\mathrm{d}}{\mathrm{d}t}\mathbb{E}[Z_t] + \frac{b}{\tau - t}\mathbb{E}[Z_t] \quad \text{and} \quad c_t^2 = \frac{\tau - t}{b}\frac{1}{\mathbb{E}[Z_t]}\left(\frac{\mathrm{d}}{\mathrm{d}t}\mathbb{V}[Z_t] + \frac{2b}{\tau - t}\mathbb{V}[Z_t]\right), \quad 0 \le t \le \tau, \tag{41}$$

where $\mathbb{E}[Z_t]$ and variance $\mathbb{V}[Z_t]$ are assumed to be modelled as

$$\mathbb{E}[Z_t] = p_0\left(\frac{t}{\tau}\right)^{p_1}\left(1 - \frac{t}{\tau}\right)^{p_2} \quad \text{and} \quad \mathbb{V}[Z_t] = q_0\left(\frac{t}{\tau}\right)^{q_1}\left(1 - \frac{t}{\tau}\right)^{q_2}, \quad 0 \le t \le \tau \tag{42}$$

with shape parameters $p_1, p_2, q_1, q_2 > 0$ and scaling parameters $p_0, q_0 > 0$. There are no data about unit-time fish counts of *P. altivelis* in the Hii River system; however, statistical analysis of 10-min fish count data of the fish in another river in Jaspan suggested that the shape parameters $p_1, p_2$ are approximately 5 to 10 and that the coefficient of variation $\sqrt{\mathbb{V}[Z_t]} / \mathbb{E}[Z_t]$ is approximately 2 [48]. These observations combined with the assumption that the eDNA shedding $G$ is proportional to the number of migrants suggested to set $q_0 = 4p_0^2$ and $q_i = 2p_i = 20$ ($i = 1, 2$), as in Yoshioka and Louriki [49]. Because the parameter $p_0$ can be absorbed into $G$, we set $p_0 = 10^6$ [49].

We also estimate parameter values in (3). The time-average flow speeds $u$ at Kisuki and Shin-Mitoya in spring to summer (March 14 to August 22) were estimated to be 0.654 (m/s) and 1.14 (m/s), respectively, based on the observation data (U-20, Onset, placed around the two sites combined with the public water level data[1]). For the ground speed $v$, it has been estimated that the migration speed of *P. altivelis* is 0.3 to 0.5 (km/day) for rivers with high anthropogenic pressure and 2 to 3 (km/day) for more natural river environments[2]. Considering that the Hii River has a mix of natural and artificial river environments, we use the intermediate value 1 (km/day) or equivalently 1/86.4 (m/s). Alternatively, one may consider using the analytical formula between the flow speed and swimming speed of *P. altivelis* [79];

---

[1] Water Information System: https://www1.river.go.jp/. Last accessed on July 31, 2026.

[2] Takahashi Research Office of Freshwater Biology: https://hito-ayu.net/introduction01.html. Last accessed on July 22, 2026.

however, simply using this in our model does not work because this formula does not account for rest events, such as meals and sleep. Indeed, if we simply apply the formula to our case, fish migrate more than 10 (km) per day, which is unrealistically long. Introducing a regime switching mechanism between rest and swim would resolve this issue but yields a far more complicated model and is therefore not discussed in this paper.

At each site, we consider the observed eDNA concentration data as $Y_t(0)$. Because we do not have high-frequency or multiple sample paths, we estimate $g_t$ and $r_t$ as constants. Based on a naive least-squares fitting between the observed eDNA concentration data and the average $\mathbb{E}\left[Y_t(0)\right]$, which is given in a closed form as follows. By approximating $L=+\infty$, the method of characteristics shows

$$\mathbb{E}\left[Y_t(x)\right]=g\int_0^t e^{-r(t-s)}\mathbb{E}\left[Z_{s-(x+u(t-s))/v}\right]\mathrm{d}s, \tag{43}$$

where the expectation in (43) can be obtained from (42). Here, the parameters to be estimated are the initial time, $g$ for source, and $r$ for decay. Finally, the parameter estimation is conducted by using the observed concentration data before July 14 considering $\tau=127$ (day). Note that, at least, the mean measure $\mathbb{E}\left[\mu_t(\mathrm{d}x)\right]$ admits a density, because taking the expectation of (9) yields a linear transport equation with an absolutely continuous source term.

We set $L=15$ (km) considering that there is a large dam called Obara Dam approximately 16 (km) upstream from Kisuki and the bank of its downstream reach is reinforced with concrete. We set $L=30$ (km) for the Shin-Mitoya case because the length from Shin-Mitoya to the source of the Mitoya River is approximately 34 to 35 (km) but empirically there will be no *P. altivelis* near the headwater area. Note that using larger values of $L$ for the Shin-Mitoya case does not critically affect the results obtained in this paper. We use the following computational resolution, which has preliminarily been found to be sufficiently fine for the analysis in this paper: $h=0.025$ (km) and $\rho=127/8000=0.015875$ (day). The unit of $\sigma$ is copies$^{1/2}$ m$^{1/2}$/ml$^{1/2}$/s$^{1/2}$, and we set $\sigma=0.5$ unless otherwise specified. The choice of $\sigma$ will be discussed later.

The computation time for one sample path is a couple of seconds for both Kisuki and Sin-Mitoya by using common laptop PCs without any parallelization. We use 100,000 samples to compute the statistics of the eDNA concentration. Codes are available online[3]. The estimated parameter values at Kisuki and Shin-Mitoya are summarized in **Table 1**. **Figure 4** compares the observed and fitted eDNA concentrations, demonstrating their reasonable agreement; for both cases, the model results are able to capture the peaks of the actual measurements.

[3] https://github.com/HidekazuYoshioka/SPDE_eDNA.git (last updated on August 3, 2026)

**Table 2.** Estimated parameter values at Kisuki and Shin-Mitoya.

| Parameter | Kisuki | Shin-Mitoya | Source |
|---|---|---|---|
| $p_0$ (-) | $10^6$ | | |
| $p_1$ (-) | 10 | | |
| $p_2$ (-) | 10 | | |
| $q_0$ (-) | $4\times10^{12}$ | | [49] |
| $q_1$ (-) | 20 | | |
| $q_2$ (-) | 20 | | |
| $b$ (-) | 61.9 | | |
| $u$ (km/day) | 56.5 | 98.5 | |
| $v$ (km/day) | 1 | | |
| $g$ (copies/ml/day) | 1453.6 | 6546.5 | This study |
| $r$ (1/day) | 5.70 | 10.3 | |
| Initial time 0 | 2025/3/27 | 2025/3/20 | |
| $L$ (km) | 15 | 30 | |

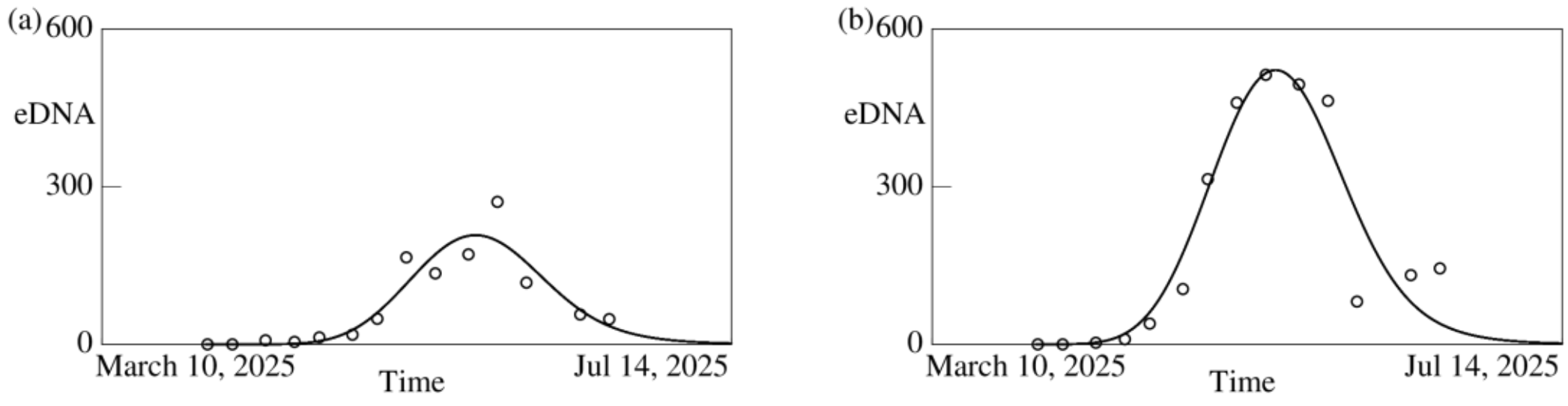

**Figure 4.** Comparison between observed (circles) and fitted eDNA concentrations (curves) in units of copies/ml from March 10 to July 14 in 2025: (a) Kisuki and (b) Shin-Mitoya.

### 4.3 Results and discussion

First, we investigate profiles of eDNA concentrations. **Figure 5** shows the computed processes $Z$ and $Y(0)$ for the Shin-Mitoya case, where the latter is approximated by the computed $Y$ at the left-most grid point (12.5 (m) from the boundary $x=0$). One may also be interested in the spatial average of eDNA concentration in a target reach to capture macroscopic fish migration dynamics. Therefore, **Figure 5** also shows a sample path of the spatial average

$$\bar{Y}_t = \frac{1}{L}\int_0^L Y_t(x)\,\mathrm{d}x,\quad 0 \le t \le T\ . \tag{44}$$

The process $Z$ is intermittent, which is due to violating the Feller condition [48]. The eDNA concentration responds to fluctuations in $Z$ with delay and attenuation, which is due to the delayed input into the SPDE model as well as its spatial structure. One finding is that the spatial average of the eDNA concentration and the value at the downstream-end do not necessarily exhibit similar fluctuation patterns, which is due to the spatial concentration profile, as shown below.

**Figure 6** shows the spatiotemporal profile of eDNA concentration for Shin-Mitoya, showing that the front of the eDNA concentration moves upstream due to the source term, while the concentration profile extends upstream due to transport. In the present model, migrants are assumed to continuously move toward the upstream-end, while in reality, they may settle down somewhere along the way. This limitation can be overcome by connecting a habitat suitability model [80] to our SPDE. The development of such a model would require a spatial survey of eDNA concentrations in fish along the river. The results obtained in this study are expected to directly serve as a foundation for such advanced research. We also report here that the finiteness of the domain length $L$ appears significant for the Kisuki case (see **Figures A1 and A2** in **Appendix**). In the rest of this paper, we therefore focus on the Shin-Mitoya case.

Second, we investigate the average (Ave) and standard deviation (Std) of the computed eDNA concentration $Y$, as shown in **Figures 7 and 8**, respectively. These cumulants have a dominant front in space that moves upstream as time elapses and gradually decreases to 0 after the migration at the downstream-end is terminated ($t > \tau$). The spatial discontinuity observed for sample paths (e.g., **Figure 6**) is averaged out in **Figures 7-8**.

Next, we investigate the influences of the noise intensity $\sigma$ on numerical solutions by comparing sample paths and statistics of the spatial average (44). **Figure 9** shows the computed average and standard deviation of the spatially averaged eDNA concentration $\bar{Y}$ for different levels of $\sigma$. **Figure 10** shows the corresponding sample paths of $\bar{Y}$. In view of sample path, the daily scale eDNA sampling for fish migration Searcy et al. [81] suggested that the eDNA concentration during migration periods is fluctuating but not intermittent. If this observation applies to the proposed model, then **Figure 10** implies that the order of $\sigma$ is at most 0.1 to 1 for this case study. The computed average of $\bar{Y}$ does not depend on $\sigma$ as expected, and the standard deviation of $\bar{Y}$ is increasing for $\sigma$ but its dependence is not explosive for $\sigma$. In contrast, as shown in **Figure 10**, fluctuations in sample paths of $Y$ at the downstream-end depend critically on $\sigma$. These computational results offer an important future perspective for SPDE-based eDNA modeling. Specifically, they suggest that collecting high-resolution eDNA data in time may enable the estimation of the $\sigma$ value, and this will not be achieved by investigating spatially averaged profiles. The results also suggest that the intensity and significance of "noise" should not be treated as equivalent between spatially averaged models and those that are not.

Finally, we investigate the extinction time of eDNA from the domain, where the extinction here is due to decay, non-Lipschitz stochasticity (Proposition 1.2.15 in Alfonsi [82]), or free outflow from $D$ after the source vanishes ($t > \tau$). We set the following extinction time, which is a pathwise quantity:

$$\tau_{\text{ext}} = \inf\left\{t \middle| t > \tau,\ Y_t(x) = 0 \text{ for all } x \in D\right\}. \tag{45}$$

In the computation, we replace "$Y_t(x) = 0$ for all $x \in D$" by "$Y_t(x_i) < 10^{-13}$ for all $i \in I$" at each time step. **Table 3** shows the computed average and standard deviation of $\tau_{\text{ext}} / \tau$, suggesting that their quantitative dependence on $\sigma$ is not monotone but seems not to be critical. These results suggest that, in the present application study, eDNA extinction is primarily by free outflow and decay. Theoretically clarifying this phenomenon needs further mathematical analysis, which will be addressed in the future based on pathwise investigations of the SPDE model.

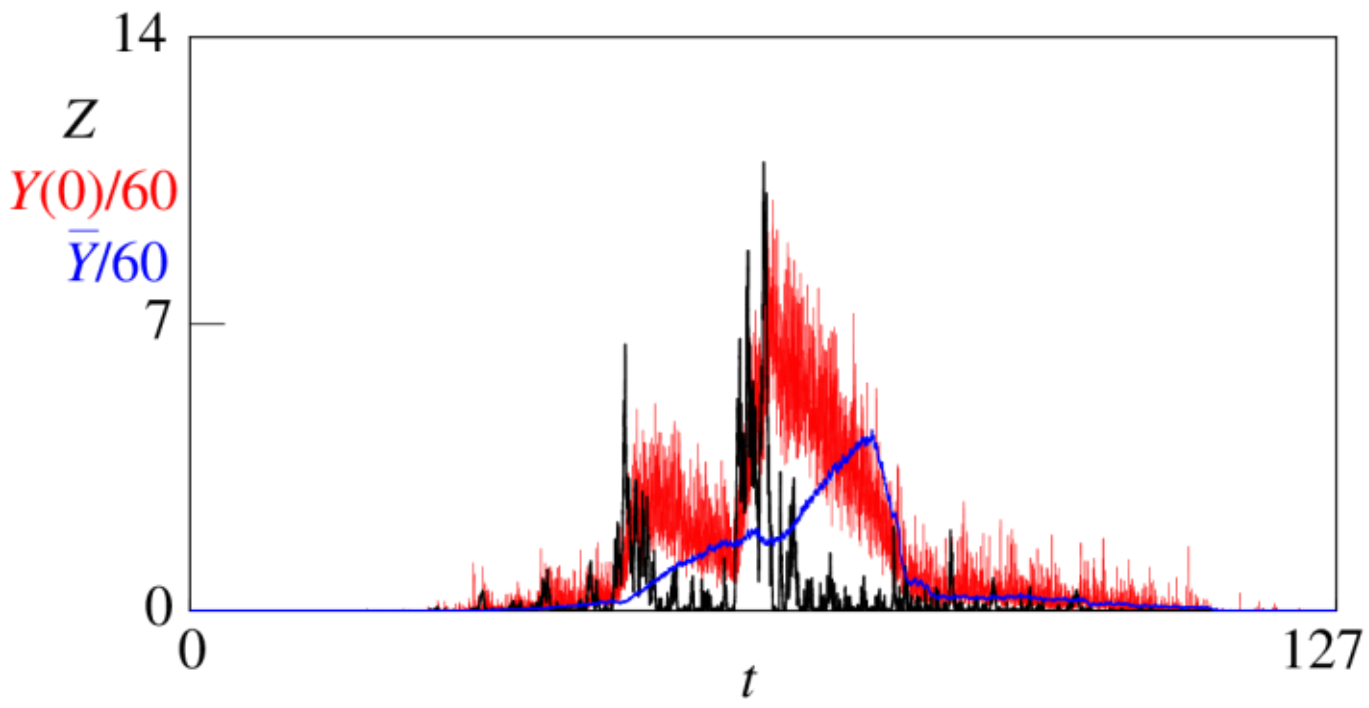

**Figure 5**. Computed processes $Z$ (-), $Y(0)$ (copies/ml), and the spatial average $\bar{Y}$ (copies/ml), for the Shin-Mitoya case, where the latter is approximated by the computed $Y$ at the left-most grid point.

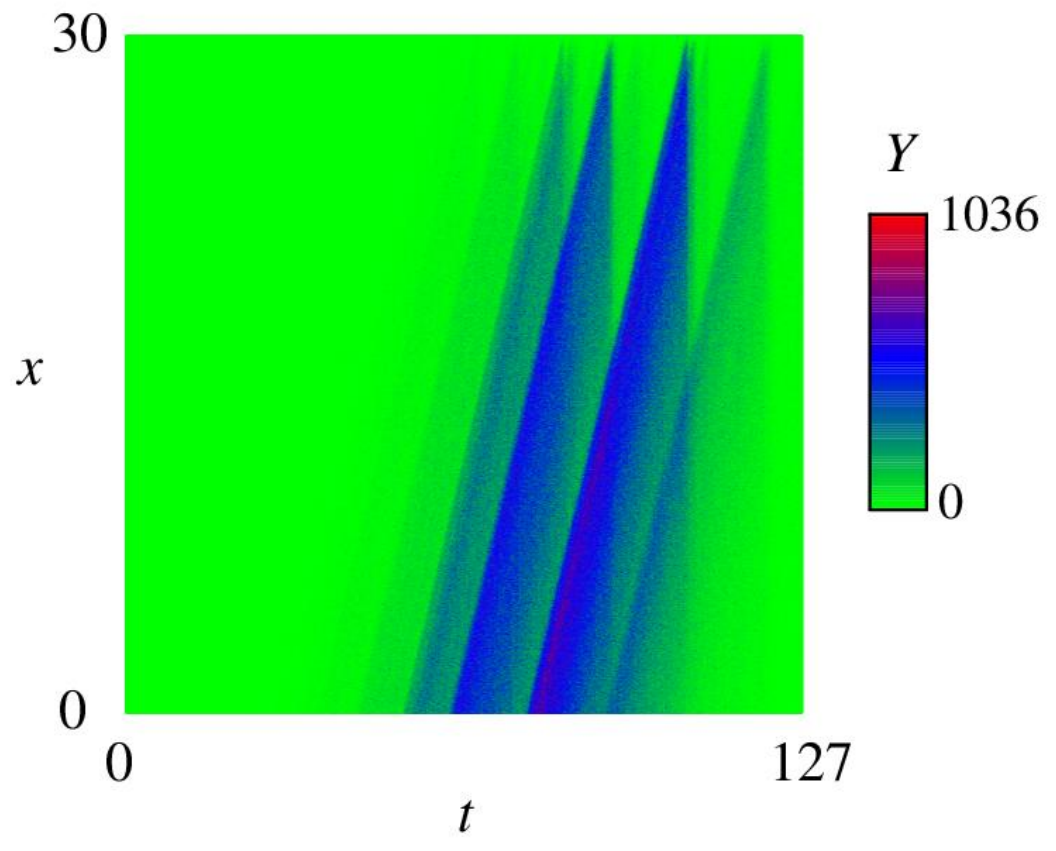

**Figure 6.** Computed sample path of the eDNA concentration $Y$ (copies/ml) for the Shin-Mitoya case at selected time instances.

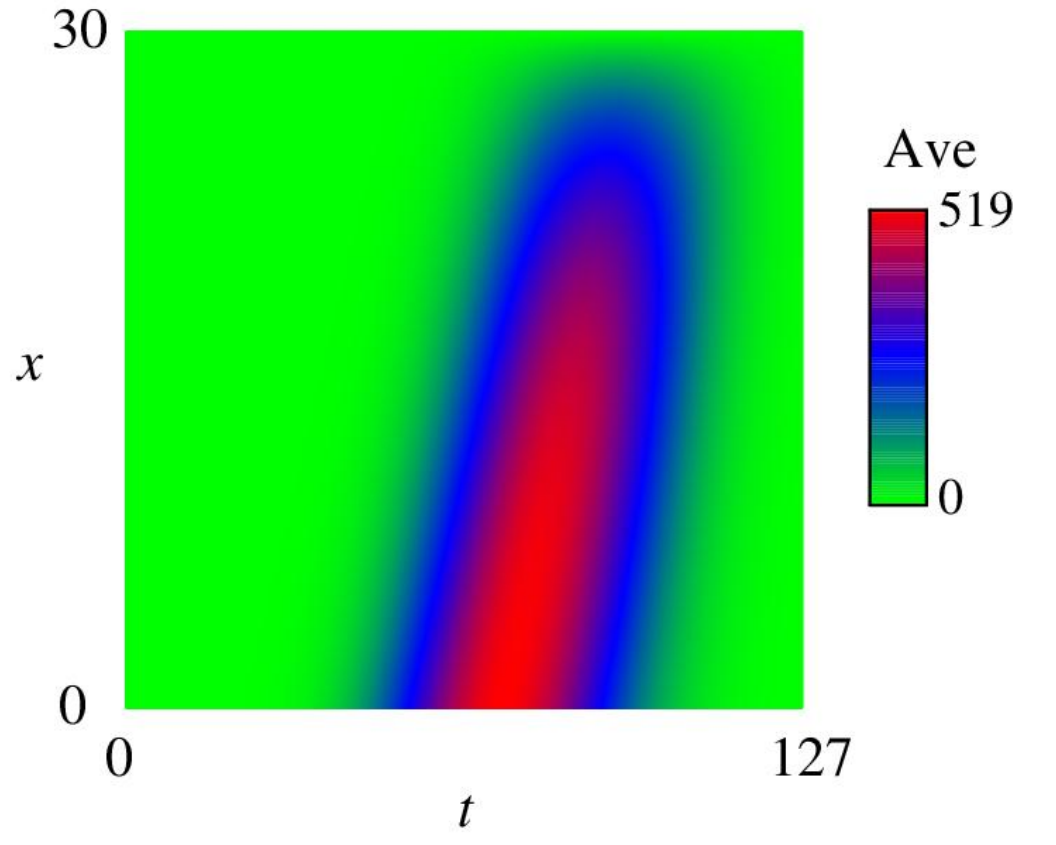

**Figure 7.** Computed average (Ave) (copies/ml) of the eDNA concentration.

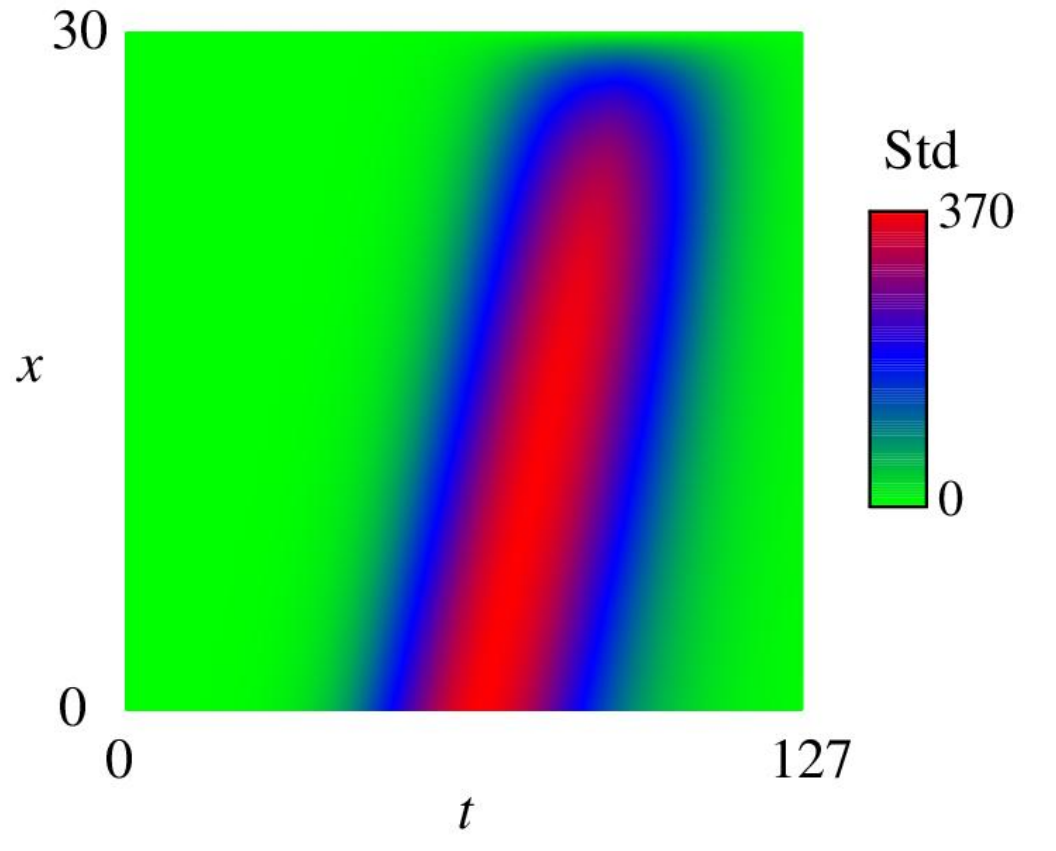

**Figure 8.** Computed standard deviation (Std) (copies/ml) of the eDNA concentration.

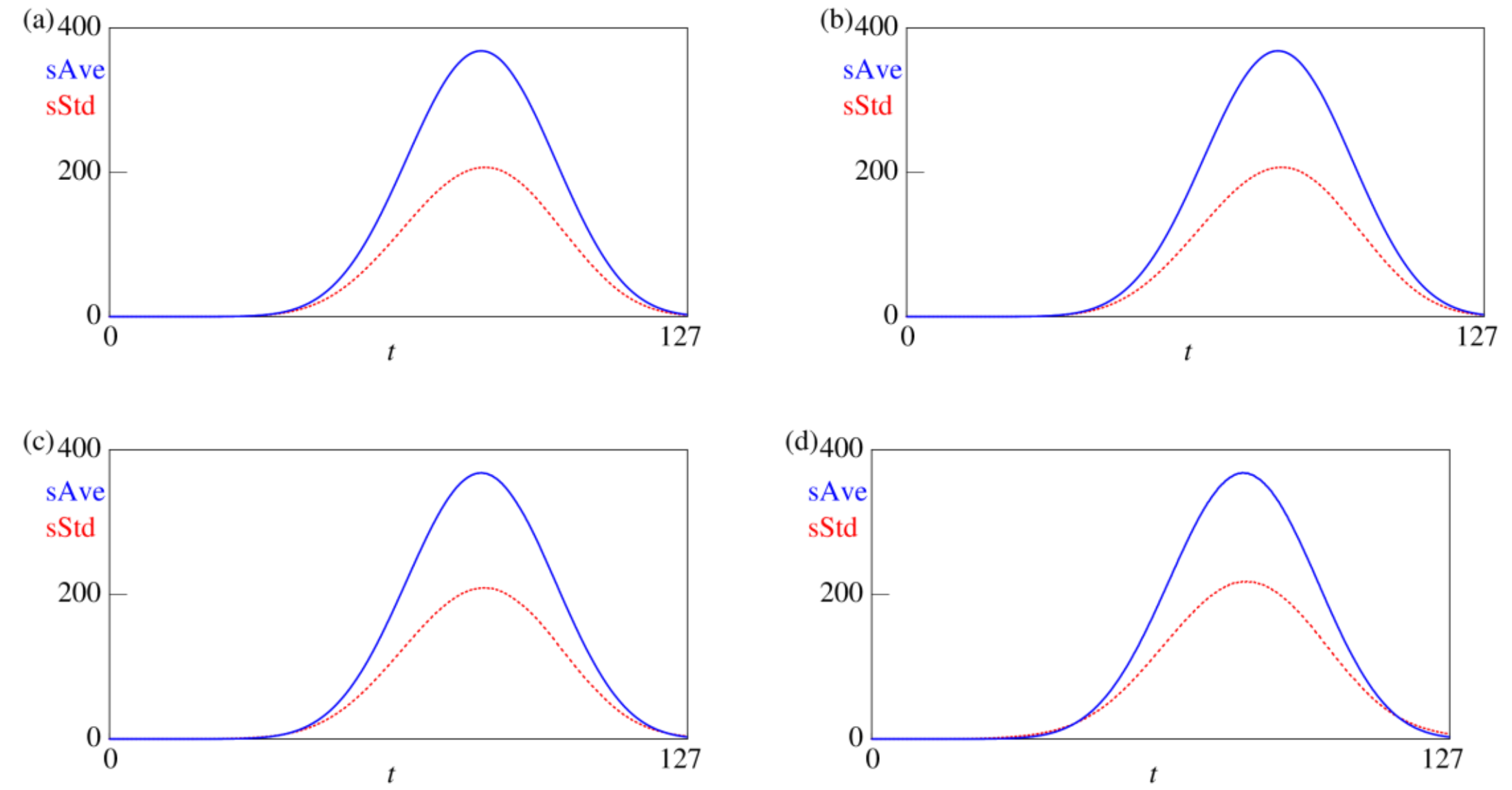

**Figure 9.** Computed average (sAve, blue solid curve) and standard deviation (sStd, red dashed curve) of the spatially averaged eDNA concentration $\bar{Y}$ (copies/ml): (a) $\sigma = 0.5$, (b) $\sigma = 1$, (c) $\sigma = 5$, (d) $\sigma = 10$.

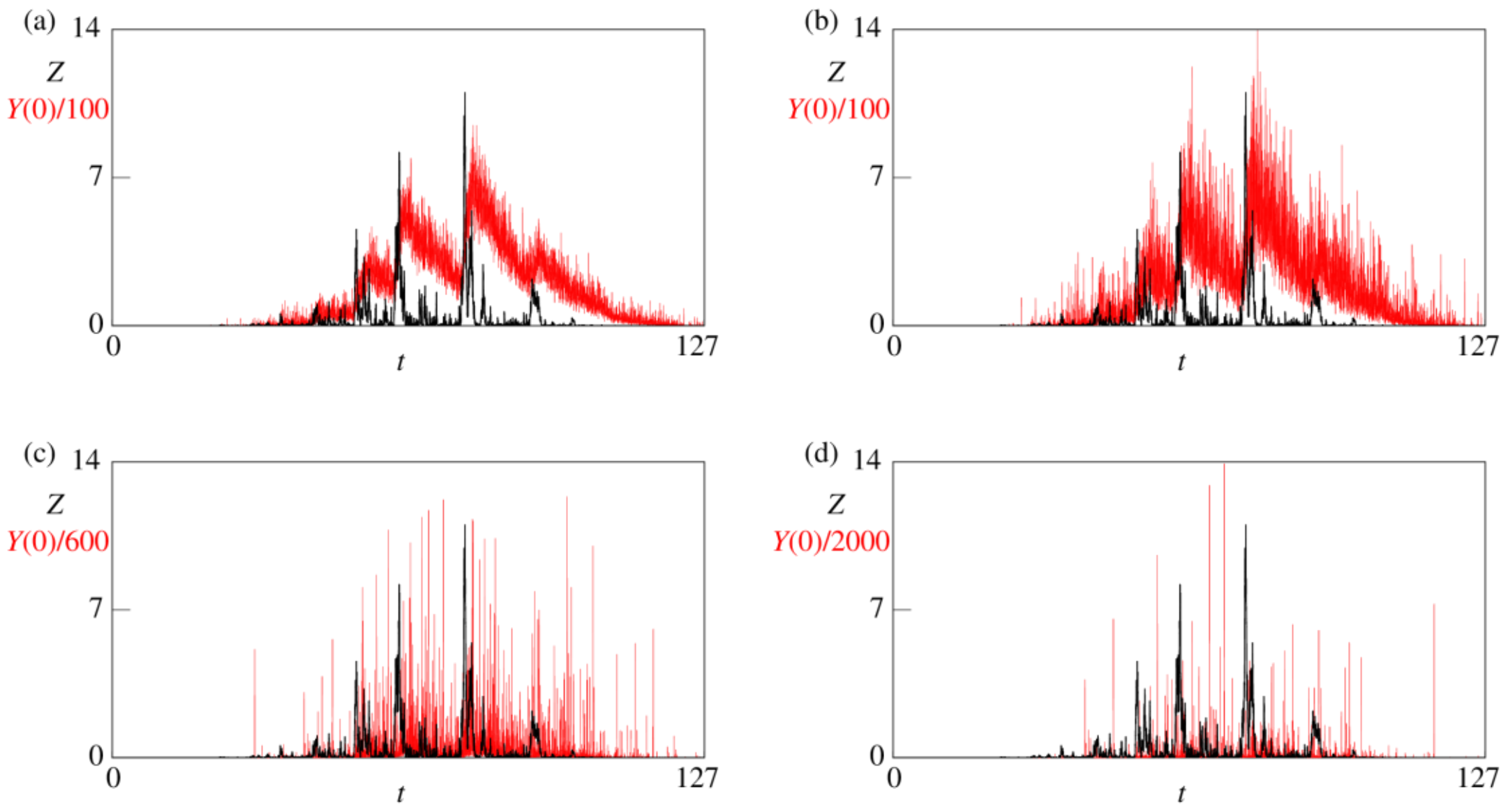

**Figure 10.** Computed sample paths of $Z$ (-) (black) and eDNA concentration $Y$ (copies/ml) (red) at the downstream-end: (a) $\sigma = 0.5$, (b) $\sigma = 1$, (c) $\sigma = 5$, (d) $\sigma = 10$. Note the difference in the vertical scale.

**Table 3.** Computed average and standard deviation of $\tau_{\text{ext}}$ for different values of $\sigma > 0$.

| $\sigma$ | Average of $\tau_{\text{ext}} / \tau$ (-) | Standard deviation of $\tau_{\text{ext}} / \tau$ (-) |
|---|---|---|
| 0.1 | 1.205 | 0.001950 |
| 0.5 | 1.201 | 0.002435 |
| 1 | 1.200 | 0.003063 |
| 5 | 1.226 | 0.000544 |
| 10 | 1.226 | 0.000595 |
| 20 | 1.226 | 0.000592 |

## 5. Conclusions

We proposed an SPDE model for spatiotemporal eDNA dynamics of migrating fish in river environments. The affine nature of the model guarantees the nonnegativity of its solution and the derivation of closed-form Laplace functional. A semidiscrete version of the model was also presented, which served as a building block for theoretical analysis as well as a computational method to generate nonnegative numerical solutions. An application case study of *P. altivelis* in the Hii River system demonstrated how the proposed model behaves and how it can be used to analyze spatiotemporal eDNA dynamics.

This study is a first step toward eDNA dynamics modeling, and many challenges remain to be addressed. For example, a limitation of this study is that the proposed model describes only eDNA dynamics along 1-D rivers, while actual fish migration phenomena occur in river networks. The domain of SPDEs in such a case would be a graph-node network where a proper internal boundary condition is needed at each node. The semidiscrete formulation for SPDEs also applies to this advanced case if the drift and diffusion coefficients as well as the internal boundary conditions satisfy the affine nature. A mathematical challenge in such cases is the well-posedness of the corresponding SPDE model. Open issues regarding the proposed model are whether its solution fully admits a density and possibility of strong convergence. These issues will be more difficult to address for higher-dimensional cases.

We did not address inverse problems inferring the fish population from eDNA data, which is an important but truly difficult problem. In fact, in this paper, we somehow identified and simulated eDNA concentrations based on assumptions regarding fish population dynamics. However, the corresponding inverse problem, specifically, the inverse problem for SPDEs, will likely prove to be far from straightforward. We are currently formulating a plan to potentially resolve this in the future, in conjunction with field surveys (e.g., longitudinal sampling) on fish population dynamics, which will be necessary to complete the inference problem. If finer data collection of spatiotemporal eDNA concentration profiles becomes possible, then the coefficients in the proposed SPDE model can be better identified. For example, one may become able to identify the intensity and correlation of noise, potentially improving the parameter identification procedure used in this paper. It is also important to conduct field surveys in collaboration with industries such as fisheries, as in this study, perform mathematical analyses based on them, and return the results back to the industry. We will continue this kind of industry-academia collaborative research based on mathematical sciences.

We are currently investigating a more realistic SPDE in which the river flow is randomized and longitudinal dispersion (i.e., a second-order spatial partial differential term) is considered. Although it is not an affine process, there is a possibility that it can be approximately regarded as an affine process. Additional difficulties in this case would be the unboundedness and discontinuity of river flow and the loss of the cascading nature of the SPDE and its discrete counterpart. Another key factor not considered in this paper is the growth of individual fish that potentially affects the shedding of eDNA; a limitation at this stage is that the existing growth models of *P. altivelis* in the Hii River system apply only from April to May (e.g., Yoshioka et al. [83]) and hence do not cover the entire migration period. In this view, a survey covering a broader time range is desired.

## Appendix

### A1. Proofs

Proofs of the propositions stated in the main text are presented. Some proofs are divided into steps because they are considered technical and long.

***Proof of Proposition 1***

**Step 1: Preliminary results about** $Z$

We have the following variation-of-constant formula for $Z$ :

$$Z_t = \begin{cases} \int_0^t a_s \left(\frac{\tau - t}{\tau - s}\right)^b \mathrm{d}s + \int_0^t c_s \left(\frac{\tau - t}{\tau - s}\right)^b \sqrt{\frac{b}{\tau - s} Z_s} \mathrm{d}B_s \ \left(= A_{Z,t} + M_{Z,t}\right) & (0 < t < \tau) \\ 0 & (t \le 0,\ t \ge \tau) \end{cases} . \tag{46}$$

By the boundedness and continuity of $a, c$, the first term $A_{Z,t}$ (deterministic term) is bounded and continuous on $[0, \tau]$, and the second term $M_{Z,t}$ (fluctuation term) satisfies (e.g., Proposition 2 in Yoshioka [48])

$$\mathbb{E}\left[M_{Z,t}\right] = 0,\ \mathbb{E}\left[\left(M_{Z,t}\right)^2\right] \to 0 \ \text{ as } \ t \nearrow \tau , \tag{47}$$

implying the convergence $M_{Z,t} \underset{t \to \tau}{\to} 0$ in the sense of mean square. Moreover, inspecting the form of (46) (Eqs. (31)-(37) and Section A.2 in Yoshioka [46] for specific cases) shows that there are some constants $C_Z > 0$ and $\kappa > 0$ such that

$$\mathbb{V}\left[Z_t\right] = \mathbb{E}\left[\left(M_{Z,t}\right)^2\right] \le C_Z \left(\tau - t\right)^{\kappa},\ \ 0 < t < \tau . \tag{48}$$

These observations lead to $M_{Z,t} \to 0$ as $t \nearrow \tau$ with probability 1, showing the global continuity of $Z$ (see **Lemma A1 in Section A3**, which technically uses Markov's inequality (Theorem 5.11 in Klenke [84] and Borel–Cantelli lemma (Theorem 2.7 in Klenke [84])).

**Step 2: semidiscrete model for** $i = N$

As in p.5 in Alfonsi [82], the following SDE subject to the initial condition $Y_0^{(N)} = 0$ admits a unique pathwise solution that is continuous with probability 1 (see also Corollary 3.23 and Theorem 4.22 in Chapter 5 of Karatzas and Shreve [85] referred in Alfonsi [82]):

$$\mathrm{d}Y_t^{(N)} = \left(g_t^{(N)} Z_{t-\tau_N} - \left(\frac{u}{h} + r_t^{(N)}\right) Y_t^{(N)}\right) \mathrm{d}t + \frac{\sigma_t^{(N)}}{\sqrt{h}} \sqrt{\left|Y_t^{(N)}\right|} \mathrm{d}W_t^{(N)},\ \ t > 0 . \tag{49}$$

The nonnegativity of its solution follows from the argument analogous to that in p.5-6 in Alfonsi [82]; the original proof is for a constant-coefficient version of (49), but the present variable coefficient case can be dealt with by noting that the coefficients $g_t^{(N)} Z_{t-\tau_N}, \frac{u}{h} + r_t^{(N)}, \frac{\sigma_t^{(N)}}{\sqrt{h}}$ are bounded, continuous, and nonnegative with probability 1, and $\mathbb{E}\left[Z_{t-\tau_N}\right] < +\infty$ . These observations show that the SDE (4) subject to the initial condition $Y_0^{(N)} = 0$ admits a unique pathwise solution that is nonnegative and continuous with probability 1.

**Step 3:** $i < N$

Cases $i < N$ are inductively proven by noting that the SDE (4) is rewritten as

$$\mathrm{d}Y_t^{(i)} = \left( \frac{u}{h} Y_t^{(i+1)} + g_t^{(i)} Z_{t-\tau_i} - \left( \frac{u}{h} + r_t^{(i)} \right) Y_t^{(i)} \right) \mathrm{d}t + \frac{\sigma_t^{(i)}}{\sqrt{h}} \sqrt{Y_t^{(i)}} \mathrm{d}W_t^{(i)}, \quad t > 0, \quad i = 1,2,3,...,N-1 \tag{50}$$

subject to the initial condition $Y_0^{(i)} = 0$. By induction, if $\left( Y_t^{(i+1)} \right)_{t \geq 0}$ uniquely exists pathwise and is nonnegative and continuous, then, as for the case $i = N$, we obtain the unique existence of $\left( Y_t^{(i)} \right)_{t \geq 0}$ and its nonnegativity. We can iterate this procedure toward $i = 1$.

□

***Proof of Proposition 2***

The proof follows the same procedure as in Proof of Proposition 3 in Guinea Juliá and Caro-Carretero [63]. We first assume $\tau > T$, and after that, we address the case $\tau \leq T$.

The necessity of the functional form of $C_N(t,T)$ is due to the linear dependence of the drift on $Y_t^{(i)}$ and $Z_{t-\tau_i}$ along with the fact that there is fixed delay $\tau_i$:

$$\exp\left( -h\lambda^{(i)} Y_T^{(i)} \right) = \exp \left( \begin{array}{l} -h\lambda^{(i)} Y_t^{(i)} \\ -h\lambda^{(i)} \int_t^T \left( \left( \frac{u}{h} Y_s^{(i+1)} + g_s^{(i)} Z_{s-\tau_i} - \left( \frac{u}{h} + r_t^{(i)} \right) Y_s^{(i)} \right) \mathrm{d}t + \frac{\sigma_s^{(i)}}{\sqrt{h}} \sqrt{Y_s^{(i)}} \mathrm{d}W_s^{(i)} \right) \mathrm{d}s \end{array} \right). \tag{51}$$

for $0 \leq t \leq T$. This implies that $C_N(t,T)$ needs to have an exponential affine form like (12), i.e., the quantity inside "exp" should be affine with respect to $Z_t$, $Y_t^{(i)}$, $\left( Z_s \right)_{t-\tau_i \leq s \leq t}$ under the conditional expectation $\mathbb{E}\left[ \cdot | \mathcal{F}_t \right]$.

**Step 1: Derivation**

Assume that $\tau > T$. We rewrite (12) as

$$C_N(t,T) = \exp\left( M_N(t,T) \right), \quad 0 \leq t \leq T. \tag{52}$$

Based on (12), applying Itô's formula to $C_N(t,T)$ yields

$$\mathrm{d}C_N(t,T) = e^{M_N(t,T)} \left( \mathrm{d}M_N(t,T) + \frac{1}{2} \left( \mathrm{d}M_N(t,T) \right)^2 \right), \tag{53}$$

where

$$\mathrm{d}M_N(t,T) = \left( \frac{\mathrm{d}\alpha_{N,t}}{\mathrm{d}t} + \frac{\mathrm{d}\beta_{N,t}}{\mathrm{d}t} Z_t + h \sum_{i=1}^{N} \frac{\mathrm{d}\gamma_{N,t}^{(i)}}{\mathrm{d}t} Y_t^{(i)} \right) \mathrm{d}t + \beta_{N,t} \mathrm{d}Z_t + h \sum_{i=1}^{N} \gamma_{N,t}^{(i)} \mathrm{d}Y_t^{(i)} + h \mathrm{d}\left( \sum_{i=1}^{N} \int_{t-\tau_i}^{t} \omega_{N,s,t}^{(i)} Z_s \mathrm{d}s \right) \tag{54}$$

and

$$\left(\mathrm{d}M_N(t,T)\right)^2 = \left(\beta_{N,t}\right)^2\left(\mathrm{d}Z_t\right)^2 + h^2\sum_{i=1}^{N}\left(\gamma_{N,t}^{(i)}\right)^2\left(\mathrm{d}Y_t^{(i)}\right)^2$$
$$= \left(\frac{b\left(c_t\right)^2}{\tau - t}\left(\beta_{N,t}\right)^2 Z_t + h\sum_{i=1}^{N}\left(\gamma_{N,t}^{(i)}\right)^2\left(\sigma_t^{(i)}\right)^2 Y_t^{(i)}\right)\mathrm{d}t. \tag{55}$$

For the last term in (54), we have

$$\mathrm{d}\int_{t-\tau_i}^{t}\omega_{N,s,t}^{(i)}Z_s\mathrm{d}s = \left(\omega_{N,t,t}^{(i)}Z_t - \omega_{N,t-\tau_i,t}^{(i)}Z_{t-\tau_i} + \int_{t-\tau_i}^{t}\frac{\partial\omega_{N,s,t}^{(i)}}{\partial t}Z_s\mathrm{d}s\right)\mathrm{d}t. \tag{56}$$

By (4) and (5), we obtain

$$\beta_{N,t}\mathrm{d}Z_t + h\sum_{i=1}^{N}\gamma_{N,t}^{(i)}\mathrm{d}Y_t^{(i)}$$
$$= \left\{\beta_{N,t}\left(a_t - \frac{b}{\tau - t}Z_t\right) + h\sum_{i=1}^{N}\gamma_{N,t}^{(i)}\left(u\frac{Y_t^{(i+1)}\mathbb{I}(i<N) - Y_t^{(i)}}{h} + g_t^{(i)}Z_{t-\tau_i} - r_t^{(i)}Y_t^{(i)}\right)\right\}\mathrm{d}t \tag{57}$$
$$+c_t\beta_{N,t}\sqrt{\frac{b}{\tau - t}Z_t}\mathrm{d}B_t + h\sum_{i=1}^{N}\gamma_{N,t}^{(i)}\sigma_t^{(i)}\sqrt{Y_t^{(i)}}\frac{\mathrm{d}W_t^{(i)}}{\sqrt{h}}.$$

Substituting (54)-(57) into (53) yields

$$\mathrm{d}C_N(t,T) = e^{M_N(t,T)}\left(\begin{array}{l}\dfrac{\mathrm{d}\alpha_{N,t}}{\mathrm{d}t} + \dfrac{\mathrm{d}\beta_{N,t}}{\mathrm{d}t}Z_t + h\displaystyle\sum_{i=1}^{N}\dfrac{\mathrm{d}\gamma_{N,t}^{(i)}}{\mathrm{d}t}Y_t^{(i)} \\ +\beta_{N,t}\left(a_t - \dfrac{b}{\tau - t}Z_t\right) + h\displaystyle\sum_{i=1}^{N}\gamma_{N,t}^{(i)}\left(u\dfrac{Y_t^{(i+1)}\mathbb{I}(i<N) - Y_t^{(i)}}{h} + g_t^{(i)}Z_{t-\tau_i} - r_t^{(i)}Y_t^{(i)}\right) \\ +h\displaystyle\sum_{i=1}^{N}\left(\omega_{N,t,t}^{(i)}Z_t - \omega_{N,t-\tau_i,t}^{(i)}Z_{t-\tau_i} + \int_{t-\tau_i}^{t}\dfrac{\partial\omega_{N,s,t}^{(i)}}{\partial t}Z_s\mathrm{d}s\right) \\ +\dfrac{1}{2}\left(\dfrac{b\left(c_t\right)^2}{\tau - t}\left(\beta_{N,t}\right)^2 Z_t + h\displaystyle\sum_{i=1}^{N}\left(\gamma_{N,t}^{(i)}\right)^2\left(\sigma_t^{(i)}\right)^2 Y_t^{(i)}\right)\end{array}\right)\mathrm{d}t \tag{58}$$
$$+\left(\text{Martingale terms}\right).$$

Here, "Martingale terms" correspond to the last line of (57).

We temporality assume that $C_N(t,T)$ is a martingale and later proves this. Then, terms multiplied by $\mathrm{d}t$ in (58) must vanish. Therefore, we obtain the following identities by matching coefficients:

**(constant terms)**
$$\frac{\mathrm{d}\alpha_{N,t}}{\mathrm{d}t} + \beta_{N,t}a_t = 0. \tag{59}$$

**($Z_t$ terms)**
$$\frac{\mathrm{d}\beta_{N,t}}{\mathrm{d}t} - \frac{b}{\tau - t}\beta_{N,t} + \frac{1}{2}\frac{b\left(c_t\right)^2}{\tau - t}\left(\beta_{N,t}\right)^2 + h\sum_{i=1}^{N}\omega_{N,t,t}^{(i)} = 0,\quad t<\tau. \tag{60}$$

**($Y_t^{(i)}$ terms: $i>1$)**
$$h\frac{\mathrm{d}\gamma_{N,t}^{(i)}}{\mathrm{d}t} - u\left(\gamma_{N,t}^{(i)} - \gamma_{N,t}^{(i-1)}\right) - hr_t^{(i)}\gamma_{N,t}^{(i)} + h\frac{1}{2}\left(\gamma_{N,t}^{(i)}\right)^2\left(\sigma_t^{(i)}\right)^2 = 0. \tag{61}$$

**($Y_t^{(i)}$ terms: $i=1$)**
$$h\frac{\mathrm{d}\gamma_{N,t}^{(1)}}{\mathrm{d}t} - u\gamma_{N,t}^{(1)} - hr_t^{(1)}\gamma_{N,t}^{(1)} + h\frac{1}{2}\left(\gamma_{N,t}^{(1)}\right)^2\left(\sigma_t^{(1)}\right)^2 = 0. \tag{62}$$

**($Z_{t-\tau_i}$ terms)** $$\gamma_{N,t}^{(i)} g_t^{(i)} - \omega_{N,t-\tau_i,t}^{(i)} = 0 . \tag{63}$$

**(Integral terms)** $$\frac{\partial \omega_{N,s,t}^{(i)}}{\partial t} = 0 \quad \text{for} \quad t - \tau_i < s < t , \quad i \in I . \tag{64}$$

We also obtain the terminal condition

$$\gamma_{N,T}^{(i)} = -\lambda^{(i)} \le 0 \quad \text{for} \quad i \in I , \quad \alpha_{N,T} = 0 , \quad \beta_{N,T} = 0 , \quad \omega_{N,T,T}^{(i)} = 0 \quad \text{and} \quad i \in I . \tag{65}$$

**Step 2: Solve the system**

The next task is to solve the system (59) through (64) with the terminal condition (65). The system of ODEs (61)-(62) can be solved from $i = 1$ to $i = N$. Indeed, the nonpositivity of the terminal condition (18) shows that $-\lambda^{(1)} \le \gamma_{N,t}^{(1)} \le 0$ for $0 \le t \le T$, and $\gamma_{N,t}^{(1)}$ is increasing toward 0 because $\bar{\gamma}_{N,t}^{(1)} = -\gamma_{N,T-t}^{(1)}$ satisfies the ODE that should be solved forward in time from $\bar{\gamma}_{N,0}^{(1)} = \lambda^{(1)}$:

$$h \frac{\mathrm{d}\bar{\gamma}_{N,t}^{(1)}}{\mathrm{d}t} = -u\bar{\gamma}_{N,t}^{(1)} - hr_{T-t}^{(1)} \bar{\gamma}_{N,t}^{(1)} - h\frac{1}{2}\left(\sigma_{T-t}^{(1)}\right)^2 \left(\bar{\gamma}_{N,t}^{(1)}\right)^2 \le 0 \tag{66}$$

and 0 is its equilibrium. Afterward, we can recursively obtain $-\infty < \gamma_{N,t}^{(i)} \le 0$ for $i > 1$; indeed, by induction, for $\bar{\gamma}_{N,t}^{(i)} = -\gamma_{N,T-t}^{(i)}$, we have the ODE to be solved forward in time from $\bar{\gamma}_{N,0}^{(i)} = \lambda^{(i)} \ge 0$:

$$h \frac{\mathrm{d}\bar{\gamma}_{N,t}^{(i)}}{\mathrm{d}t} = u\bar{\gamma}_{N,t}^{(i-1)} - u\bar{\gamma}_{N,t}^{(i)} - hr_{T-t}^{(i)} \bar{\gamma}_{N,t}^{(i)} - h\frac{1}{2}\left(\sigma_{T-t}^{(i)}\right)^2 \left(\gamma_{N,t}^{(i)}\right)^2 \tag{67}$$

whose solution is bounded if $\bar{\gamma}_{N,t}^{(i-1)}$ does.

By (63), we obtain

$$\omega_{N,t-\tau_i,t}^{(i)} = \gamma_{N,t}^{(i)} g_t^{(i)} , \quad t < T \tag{68}$$

and hence

$$\omega_{N,t,t+\tau_i}^{(i)} = \gamma_{N,t+\tau_i}^{(i)} g_{t+\tau_i}^{(i)} , \quad t < T - \tau_i . \tag{69}$$

By (64), we have

$$\omega_{N,s,t}^{(i)} = w_{N,s}^{(i)} \quad \text{for} \quad t - \tau_i \le s < t \tag{70}$$

with some $t$-independent coefficient $w_{N,s}^{(i)}$. This combined with the terminal condition (65) yields

$$\omega_{N,t,t}^{(i)} = 0 \quad \text{for} \quad T - \tau_i \le t < T . \tag{71}$$

Consequently, we have

$$\omega_{N,t,t}^{(i)} = \begin{cases} 0 & \left(T - \tau_i \le t < T\right) \\ \gamma_{N,t+\tau_i}^{(i)} g_{t+\tau_i}^{(i)} & \left(t < T - \tau_i\right) \end{cases} , \quad i \in I . \tag{72}$$

Afterward, we substitute (72) into (60), from which we can find $\beta_{N,t}$ because of the boundedness of $\omega_{N,t,t}^{(i)}$ and the sign of each term of this ODE. Finally, $\alpha_{N,t}$ is obtained from (59) by using $\beta_{N,t}$.

**Step 3: Remaining case**

Now, assume that $\tau \le T$. The difference from case $\tau > T$ is that we need to use the following trivial SDE for $Z$ due to $Z_t = 0$ ($t > \tau$):

$$\mathrm{d}Z_t = 0 \cdot \mathrm{d}t + 0 \cdot \mathrm{d}B_t, \quad t > \tau. \tag{73}$$

For $\tau < t < T$, we obtain

$$\mathrm{d}C_N(t,T) = e^{M_N(t,T)} \left( \begin{array}{l} \frac{\mathrm{d}\alpha_{N,t}}{\mathrm{d}t} + h\sum_{i=1}^{N} \frac{\mathrm{d}\gamma_{N,t}^{(i)}}{\mathrm{d}t} Y_t^{(i)} \\ +h\sum_{i=1}^{N} \gamma_{N,t}^{(i)} \left( u \frac{Y_t^{(i+1)} \mathbb{I}(i < N) - Y_t^{(i)}}{h} + g_t^{(i)} Z_{t-\tau_i} - r_t^{(i)} Y_t^{(i)} \right) \\ +h\sum_{i=1}^{N} \left( -\omega_{N,t-\tau_i,t}^{(i)} Z_{t-\tau_i} + \int_{\min\{t-\tau_i,\tau\}}^{\tau} \frac{\mathrm{d}\omega_{N,s,t}^{(i)}}{\mathrm{d}t} Z_s \mathrm{d}s \right) + h\sum_{i=1}^{N} \left(\gamma_{N,t}^{(i)}\right)^2 \left(\sigma_t^{(i)}\right)^2 Y_t^{(i)} \end{array} \right) \mathrm{d}t \tag{74}$$
$$+ (\text{Martingale terms}),$$

and hence we can set $\alpha_{N,t} = \beta_{N,t} = 0$ for $\tau < t < T$ from (65). We focus on some $i \in I$. If $\tau > t - \tau_i$, then $\frac{\mathrm{d}\omega_{N,s,t}^{(i)}}{\mathrm{d}t} = 0$ ($t - \tau_i < s \le \tau$) and the same holds true for $t - \tau_i \le s \le t$, and hence we obtain (72). If $\tau \ge t - \tau_i$, then the integral term involving $\frac{\mathrm{d}\omega_{N,s,t}^{(i)}}{\mathrm{d}t}$ disappears from (74), again yielding $\frac{\mathrm{d}\omega_{N,s,t}^{(i)}}{\mathrm{d}t} = 0$. Consequently, For $t < \tau$, we can solve the ODEs (60) along with (72) and afterwards (59).

From the preceding discussion, the right-hand side of (12) is bounded above by 1, $C_N(t,T)$ is a bounded local martingale and hence is a martingale.

□

***Proof of Proposition 3***

The proof proceeds in the same split as that of **Proof of Proposition 2**; we assume $\tau > T$ because the case $\tau \le T$ can be handled as in the arguments around (74). Rigorously speaking, we need temporality assume that $C(t,T)$ is a martingale and later proves this, but this follows from the nonpositivity of $\alpha, \beta, \gamma, \omega$.

**Step 1: Derivation**

Based on the ansatz that the Laplace functional has a representation on the right-hand side of (20), applying classical Itô's formula yields ($M(t,T)$ represents the quantity inside "exp" in (20))

$$
\mathrm{d}C(t,T)=e^{M(t,T)}\left(\begin{array}{l}\dfrac{\mathrm{d}\alpha_t}{\mathrm{d}t}+\dfrac{\mathrm{d}\beta_t}{\mathrm{d}t}Z_t+\displaystyle\int_0^L\dfrac{\partial\gamma_t(x)}{\partial t}\mu_t(\mathrm{d}x)\\ +\beta_t\left(a_t-\dfrac{b}{\tau-t}Z_t\right)-\displaystyle\int_0^L u\dfrac{\partial\gamma_t(x)}{\partial x}\mu_t(\mathrm{d}x)-uY_t(0)\gamma_t(0)\\ +\displaystyle\int_0^L\gamma_t(x)g_t(x)Z_{t-x/v}\mathrm{d}x-\int_0^L\gamma_t(x)r_t(x)\mu_t(\mathrm{d}x)\\ +\displaystyle\int_0^L\omega_{t,t}(x)Z_t\mathrm{d}x-\int_0^L\omega_{t-x/v,t}(x)Z_{t-x/v}\mathrm{d}x+\int_0^L\left(\int_{t-x/v}^t\dfrac{\partial\omega_{s,t}(x)}{\partial t}Z_s\mathrm{d}s\right)\mathrm{d}x\\ +\dfrac{1}{2}\left(\dfrac{b(c_t)^2}{\tau-t}(\beta_t)^2Z_t+\displaystyle\int_0^L(\gamma_t(x))^2(\sigma_t(x))^2\mu_t(\mathrm{d}x)\right)\end{array}\right)\mathrm{d}t \tag{75}
$$

$$
+(\text{Martingale terms}).
$$

Here, "Martingale terms" correspond to the terms that include the white noise increments.

Because $C(t,T)$ is a martingale by the representation (11), terms multiplied by $\mathrm{d}t$ in (75) must vanish. Therefore, we obtain the following identities by the matching coefficient approach:

**(constant terms)**
$$
\frac{\mathrm{d}\alpha_t}{\mathrm{d}t}+\beta_t a_t=0. \tag{76}
$$

**($Z_t$ terms)**
$$
\frac{\mathrm{d}\beta_t}{\mathrm{d}t}-\frac{b}{\tau-t}\beta_t+\frac{1}{2}\frac{b(c_t)^2}{\tau-t}(\beta_t)^2+\int_0^L\omega_{t,t}(x)\mathrm{d}x=0,\quad t<\tau. \tag{77}
$$

**($Y_t$ terms: $x>0$)**
$$
\frac{\partial\gamma_t(x)}{\partial t}-u\frac{\partial\gamma_t(x)}{\partial x}-r_t(x)\gamma_t(x)+\frac{1}{2}(\gamma_t(x))^2(\sigma_t(x))^2=0. \tag{78}
$$

**($Z_{t-x/v}$ terms)**
$$
\gamma_t(x)g_t(x)-\omega_{t-x/v,t}(x)=0. \tag{79}
$$

**(time integral terms)**
$$
\frac{\partial\omega_{s,t}(x)}{\partial t}=0\quad\text{for}\quad t-x/v\le s<t. \tag{80}
$$

We also obtain the boundary and terminal conditions (25) and (26). Here, the boundary condition $\gamma_t(0)=0$ comes from the term $-uY_t(0)\gamma_t(0)$ in (75).

**Step 2: Solve the system**

The next task is to solve the system (76)-(80) on the basis of boundary and terminal conditions (25) and (26). The nonlinear hyperbolic partial differential equation (78) can be solved to find a unique solution by the method of characteristics; indeed, applying the change in variables $(\varsigma,y)=(t,x-ut)$ to (78) (the range of the variable $y$ is $-uT<y<L$) yields

$$
\frac{\partial\bar{\gamma}_\varsigma(y)}{\partial\varsigma}-\bar{r}_\varsigma(y)\bar{\gamma}_\varsigma(y)+\frac{1}{2}(\bar{\gamma}_\varsigma(y))^2(\bar{\sigma}_\varsigma(y))^2=0 \tag{81}
$$

where $\bar{\gamma}_\varsigma(y)=\gamma_t(x)$, $\bar{r}_\varsigma(y)=r_t(x)$, and $\bar{\sigma}_\varsigma(y)=\sigma_t(x)$ for $0<y<L$, and $\bar{\gamma}_\varsigma(y)=0$ for $y\le 0$. For each $y$, the equation (81) is seen as an ODE for $\bar{\gamma}_\varsigma(y)$, which can be solved as in the semidiscrete case (see **Proof of Proposition 2**) from the terminal condition $\bar{\gamma}_T(y)=0$ ($y\le 0$) and $\bar{\gamma}_T(y)=-\lambda(x)$ ( $0<y<L$ ). The obtained solution for the original parameters $(t,x)$ is bounded since $-\lambda(x)\le\gamma_t(x)\le 0$; more specifically, as shown in **Proof of Proposition 4**, the nonlinear hyperbolic partial differential equation (78) has a unique entropy solution belonging to $L^1([0,T]\times D)\cap L^\infty([0,T]\times D)$ (understood weakly in time):

$$0=\int_0^L \phi(x)\left(\frac{\partial\gamma_t(x)}{\partial t}-r_t(x)\gamma_t(x)+\frac{1}{2}(\sigma_t(x)\gamma_t(x))^2\right)\mathrm{d}x+\int_0^L u\frac{\mathrm{d}\phi(x)}{\mathrm{d}x}\gamma_t(x)\mathrm{d}x-u\phi(L)\gamma_t(L) \quad (82)$$

for all $\phi\in C^\infty(\bar{D})$ (e.g., Assumption 1 and Theorem 18 in Martin [86]).

After solving (78), we obtain

$$\omega_{t,t}(x)=\gamma_{t+x/v}(x)g_{t+x/v}(x),\ \ t<T-x/v\,. \quad (83)$$

By (83), we have

$$\omega_{s,t}(x)=w_s(x)\ \text{ for }\ t-x/v\le s<t \quad (84)$$

with some $t$-independent coefficient $w_s(x)$. This combined with the terminal condition (26) yields

$$\omega_{t,t}(x)=\begin{cases}0 & (T-x/v\le t<T)\\ \gamma_{t+x/v}(x)g_{t+x/v}(x) & (t<T-x/v)\end{cases}. \quad (85)$$

Afterward, we can substitute (85) into (77), from which we can find $\beta_t$. Finally, the ODE (76) can be obtained by using this $\beta_t$.

□

***Proof of Proposition 4***

**Step 1: Several estimates about discretized** $\gamma$

For $\gamma$, we have the convergence as follows, where the proof is based on the convergence argument in LeVeque [87] with adaptations to our case. First, we have consistency of the upwind discretization:

$$\lim_{N\to+\infty}\max_{2\le i\le N}\left|\frac{\phi(x_i)-\phi(x_{i-1})}{h}-\frac{\mathrm{d}\phi(x_i)}{\mathrm{d}x}\right|\le\lim_{N\to+\infty}\max_{2\le i\le N}\frac{h}{2}\left\|\frac{\mathrm{d}^2\phi}{\mathrm{d}x^2}\right\|_\infty\to 0\ \text{ for any }\ \phi\in C^\infty(\bar{D})\,. \quad (86)$$

Moreover, we obtain the uniform boundedness result, where $\lambda^{(i)}=\lambda(x_i)$:

$$-\sup_{x\in D}\lambda(x)\le-\sup_{i\in I}\lambda^{(i)}\le\gamma_{N,t}^{(i)}\le 0 \quad (87)$$

and comparison result

$$\gamma_{1,N,t}^{(i)}\le\gamma_{2,N,t}^{(i)},\ \ 0\le t\le T\ ,\ \ i\in I\,, \quad (88)$$

where $\gamma_{k,N,t}^{(i)}$ means $\gamma_{N,t}^{(i)}$ subject to the terminal condition $-\lambda_k^{(i)}$ ( $k=1,2$ ) with $\lambda_2^{(i)}\le\lambda_1^{(i)}$ ( $i\in I$ ).

The uniform boundedness is proven as follows. For $i=1$, we have

$$\frac{\mathrm{d}\gamma_{N,t}^{(1)}}{\mathrm{d}t}=\left(\frac{u}{h}+r_t^{(1)}\right)\gamma_{N,t}^{(1)}-\frac{1}{2}\left(\sigma_t^{(i)}\right)^2\left(\gamma_{N,t}^{(i)}\right)^2=\left(\frac{u}{h}+r_t^{(1)}-\frac{1}{2}\left(\sigma_t^{(1)}\right)^2\gamma_{N,t}^{(1)}\right)\gamma_{N,t}^{(i)}\left(=Q_{1,N,t}\gamma_{N,t}^{(i)}\right). \quad (89)$$

This implies that the sign of $\gamma_{N,t}^{(1)}$ is negative or zero because 0 is an equilibrium of the ODE (89). Then, we have $Q_{1,N,t} \geq 0$. Hence, we obtain

$$0 \geq \gamma_{N,t}^{(1)} = -\lambda^{(1)} \exp\left(-\int_t^T Q_{1,N,s} \mathrm{d}s\right) \geq -\lambda^{(1)} . \tag{90}$$

For $i > 1$, if $\gamma_{N,t}^{(i-1)} \leq 0$, then

$$\frac{\mathrm{d}\gamma_{N,t}^{(i)}}{\mathrm{d}t} = \left(\frac{u}{h} + r_t^{(i)}\right)\gamma_{N,t}^{(i)} - \frac{u}{h}\gamma_{N,t}^{(i-1)} - \frac{1}{2}\left(\sigma_t^{(i)}\right)^2\left(\gamma_{N,t}^{(i)}\right)^2 \geq \left(\frac{u}{h} + r_t^{(i)} - \frac{1}{2}\left(\sigma_t^{(i)}\right)^2 \gamma_{N,t}^{(i)}\right)\gamma_{N,t}^{(i)} \left(= Q_{i,N,t}\gamma_{N,t}^{(i)}\right) \tag{91}$$

and hence by Gronwall's inequality (Theorem 8.1 in Mao [88]),

$$\gamma_{N,t}^{(i)} \leq \gamma_{N,T}^{(i)} \exp\left(-\int_t^T Q_{i,N,s} \mathrm{d}s\right) = -\lambda^{(i)} \exp\left(-\int_t^T Q_{i,N,s} \mathrm{d}s\right) \leq 0 . \tag{92}$$

Additionally, if $\gamma_{N,t}^{(i-1)} \geq -\max_{i \in I} \lambda^{(i)}$, then

$$\begin{aligned}
\frac{\mathrm{d}\left(\gamma_{N,t}^{(i)} + \max_{i \in I} \lambda^{(i)}\right)}{\mathrm{d}t} &\leq \left(\frac{u}{h} + r_t^{(i)}\right)\gamma_{N,t}^{(i)} - \frac{1}{2}\left(\sigma_t^{(i)}\right)^2\left(\gamma_{N,t}^{(i)}\right)^2 + \frac{u}{h}\max_{i \in I} \lambda^{(i)} \\
&= \frac{u}{h}\left(\gamma_{N,t}^{(i)} + \max_{i \in I} \lambda^{(i)}\right) + r_t^{(i)}\gamma_{N,t}^{(i)} - \frac{1}{2}\left(\sigma_t^{(i)}\right)^2\left(\gamma_{N,t}^{(i)}\right)^2 \\
&\leq \frac{u}{h}\left(\gamma_{N,t}^{(i)} + \max_{i \in I} \lambda^{(i)}\right),
\end{aligned} \tag{93}$$

where we used (92). By Gronwall's inequality, we obtain

$$\gamma_{N,t}^{(i)} + \max_{i \in I} \lambda^{(i)} \geq e^{-\frac{u}{h}(T-t)}\left(\gamma_{N,T}^{(i)} + \max_{i \in I} \lambda^{(i)}\right) = e^{-\frac{u}{h}(T-t)}\left(-\lambda^{(i)} + \max_{i \in I} \lambda^{(i)}\right) \geq 0 . \tag{94}$$

Therefore, by induction, we obtain the uniform boundedness (87).

The comparison result is proven as follows. Set $\Delta_{N,t}^{(i)} = \gamma_{2,N,t}^{(i)} - \gamma_{1,N,t}^{(i)}$. For $i = 1$, we have

$$\frac{\mathrm{d}\Delta_{N,t}^{(1)}}{\mathrm{d}t} = \left(\frac{u}{h} + r_t^{(1)} - \frac{1}{2}\left(\sigma_t^{(1)}\right)^2\left(\gamma_{2,N,t}^{(1)} + \gamma_{1,N,t}^{(1)}\right)\right)\Delta_{N,t}^{(1)} \quad \left(= Q'_{1,N,t}\Delta_{N,t}^{(1)}\right), \tag{95}$$

yielding

$$\gamma_{2,N,t}^{(1)} - \gamma_{1,N,t}^{(1)} = \Delta_{N,t}^{(1)} = \Delta_{N,T}^{(1)} \exp\left(-\int_t^T Q'_{1,N,s} \mathrm{d}s\right) \geq 0 \tag{96}$$

because $\Delta_{N,T}^{(1)} = \gamma_{2,N,T}^{(1)} - \gamma_{1,N,T}^{(1)} \geq 0$. For $i > 1$, assume by induction that $\Delta_{N,t}^{(i-1)} \geq 0$. Afterward, we obtain

$$\begin{aligned}
\frac{\mathrm{d}\Delta_{N,t}^{(i)}}{\mathrm{d}t} &= \left(\frac{u}{h} + r_t^{(i)} - \frac{1}{2}\left(\sigma_t^{(i)}\right)^2\left(\gamma_{2,N,t}^{(i)} + \gamma_{1,N,t}^{(i)}\right)\right)\Delta_{N,t}^{(i)} - \frac{u}{h}\Delta_{N,t}^{(i-1)} \\
&\leq \left(\frac{u}{h} + r_t^{(i)} - \frac{1}{2}\left(\sigma_t^{(i)}\right)^2\left(\gamma_{2,N,t}^{(i)} + \gamma_{1,N,t}^{(i)}\right)\right)\Delta_{N,t}^{(i)} \\
&\left(= Q'_{i,N,t}\Delta_{N,t}^{(i)}\right),
\end{aligned} \tag{97}$$

yielding the desired result

$$\gamma_{2,N,t}^{(i)} - \gamma_{1,N,t}^{(i)} = \Delta_{N,t}^{(i)} \geq \Delta_{N,T}^{(i)} \exp\left(-\int_t^T Q'_{i,N,s} \mathrm{d}s\right) \geq 0 . \tag{98}$$

From the uniform boundedness, we also obtain that the total variation is bounded:

$$\mathrm{TV}\left(\gamma_{N,t}\right)=\sum_{i=1}^{N}\left|\gamma_{N,t}^{(i)}-\gamma_{N,t}^{(i-1)}\right|\leq C',\quad 0\leq t\leq T\ , \tag{99}$$

where we set $\gamma_{N,t}^{(0)}=0$ and $C'>0$ is a constant independent of $N$. Indeed, we formally have

$$\frac{\mathrm{d}}{\mathrm{d}t}\left|\gamma_{N,t}^{(i)}-\gamma_{N,t}^{(i-1)}\right|=\mathrm{sgn}\left(\gamma_{N,t}^{(i)}-\gamma_{N,t}^{(i-1)}\right)\frac{\mathrm{d}}{\mathrm{d}t}\left(\gamma_{N,t}^{(i)}-\gamma_{N,t}^{(i-1)}\right),\quad i\in I \tag{100}$$

and

$$\mathrm{sgn}\left(\gamma_{N,t}^{(1)}-\gamma_{N,t}^{(0)}\right)\frac{\mathrm{d}}{\mathrm{d}t}\left(\gamma_{N,t}^{(1)}-\gamma_{N,t}^{(0)}\right)=-\frac{\mathrm{d}\gamma_{N,t}^{(1)}}{\mathrm{d}t}=-\frac{u}{h}\gamma_{N,t}^{(1)}-r_t^{(1)}\gamma_{N,t}^{(1)}+\frac{1}{2}\left(\sigma_t^{(1)}\right)^2\left(\gamma_{N,t}^{(1)}\right)^2\geq-\frac{u}{h}\gamma_{N,t}^{(1)} \tag{101}$$

for $i=1$ because of $\gamma_{N,t}^{(1)}\leq 0$, and

$$\begin{aligned}\frac{\mathrm{d}}{\mathrm{d}t}\left(\gamma_{N,t}^{(i)}-\gamma_{N,t}^{(i-1)}\right)&=u\frac{\gamma_{N,t}^{(i)}-2\gamma_{N,t}^{(i-1)}+\gamma_{N,t}^{(i-2)}}{h}\\&\quad+r_t^{(i)}\gamma_{N,t}^{(i)}-r_t^{(i-1)}\gamma_{N,t}^{(i-1)}-\frac{1}{2}\left(\gamma_{N,t}^{(i)}\right)^2\left(\sigma_t^{(i)}\right)^2+\frac{1}{2}\left(\gamma_{N,t}^{(i-1)}\right)^2\left(\sigma_t^{(i-1)}\right)^2\\&=u\frac{\gamma_{N,t}^{(i)}-2\gamma_{N,t}^{(i-1)}+\gamma_{N,t}^{(i-2)}}{h}\\&\quad+r_t^{(i-1)}\left(\gamma_{N,t}^{(i)}-\gamma_{N,t}^{(i-1)}\right)+\frac{1}{2}\left(\sigma_t^{(i)}\right)^2\left(\left(\gamma_{N,t}^{(i-1)}\right)^2-\left(\gamma_{N,t}^{(i)}\right)^2\right)\\&\quad+\left(r_t^{(i)}-r_t^{(i-1)}\right)\gamma_{N,t}^{(i)}-\frac{1}{2}\left(\gamma_{N,t}^{(i-1)}\right)^2\left(\left(\sigma_t^{(i)}\right)^2-\left(\sigma_t^{(i-1)}\right)^2\right)\end{aligned} \tag{102}$$

for $i>1$. Thus, we obtain

$$\begin{aligned}&\mathrm{sgn}\left(\gamma_{N,t}^{(i)}-\gamma_{N,t}^{(i-1)}\right)\left\{\begin{aligned}&r_t^{(i-1)}\left(\gamma_{N,t}^{(i)}-\gamma_{N,t}^{(i-1)}\right)+\frac{1}{2}\left(\sigma_t^{(i)}\right)^2\left(\left(\gamma_{N,t}^{(i-1)}\right)^2-\left(\gamma_{N,t}^{(i)}\right)^2\right)\\&+\left(r_t^{(i)}-r_t^{(i-1)}\right)\gamma_{N,t}^{(i)}-\frac{1}{2}\left(\gamma_{N,t}^{(i-1)}\right)^2\left(\left(\sigma_t^{(i)}\right)^2-\left(\sigma_t^{(i-1)}\right)^2\right)\end{aligned}\right\}\\&\geq-C_{r,\sigma}h\left|\gamma_{N,t}^{(i)}\right|+\mathrm{sgn}\left(\gamma_{N,t}^{(i)}-\gamma_{N,t}^{(i-1)}\right)\left\{r_t^{(i-1)}\left(\gamma_{N,t}^{(i)}-\gamma_{N,t}^{(i-1)}\right)-\frac{1}{2}\left(\sigma_t^{(i)}\right)^2\left(\left(\gamma_{N,t}^{(i)}\right)^2-\left(\gamma_{N,t}^{(i-1)}\right)^2\right)\right\}\\&\geq-C_{r,\sigma}h\left\|\lambda\right\|_\infty+\left(r_t^{(i-1)}-\frac{1}{2}\left(\sigma_t^{(i)}\right)^2\left(\gamma_{N,t}^{(i)}+\gamma_{N,t}^{(i-1)}\right)\right)\mathrm{sgn}\left(\gamma_{N,t}^{(i)}-\gamma_{N,t}^{(i-1)}\right)\left(\gamma_{N,t}^{(i)}-\gamma_{N,t}^{(i-1)}\right)\\&=-C_{r,\sigma}h\left\|\lambda\right\|_\infty+\left(r_t^{(i-1)}-\frac{1}{2}\left(\sigma_t^{(i)}\right)^2\left(\gamma_{N,t}^{(i)}+\gamma_{N,t}^{(i-1)}\right)\right)\left|\gamma_{N,t}^{(i)}-\gamma_{N,t}^{(i-1)}\right|\\&\geq-C_{r,\sigma}h\left\|\lambda\right\|_\infty\end{aligned} \tag{103}$$

with a constant $C_{r,\sigma}>0$ depending only on $r,\sigma$. We also have the following estimate due to $-\gamma_{N,t}^{(1)}=\left|\gamma_{N,t}^{(1)}\right|$ and $\gamma_{N,t}^{(0)}=0$:

$$
\begin{aligned}
&\sum_{i=2}^{N}\operatorname{sgn}\left(\gamma_{N,t}^{(i)}-\gamma_{N,t}^{(i-1)}\right)\left(\gamma_{N,t}^{(i)}-2\gamma_{N,t}^{(i-1)}+\gamma_{N,t}^{(i-2)}\right)-\gamma_{N,t}^{(1)}\\
&=\sum_{i=2}^{N}\left|\gamma_{N,t}^{(i)}-\gamma_{N,t}^{(i-1)}\right|-\sum_{i=2}^{N}\operatorname{sgn}\left(\gamma_{N,t}^{(i)}-\gamma_{N,t}^{(i-1)}\right)\left(\gamma_{N,t}^{(i-1)}-\gamma_{N,t}^{(i-2)}\right)-\gamma_{N,t}^{(1)}\\
&=\sum_{i=1}^{N}\left|\gamma_{N,t}^{(i)}-\gamma_{N,t}^{(i-1)}\right|-\sum_{i=2}^{N}\operatorname{sgn}\left(\gamma_{N,t}^{(i)}-\gamma_{N,t}^{(i-1)}\right)\left(\gamma_{N,t}^{(i-1)}-\gamma_{N,t}^{(i-2)}\right)\\
&=\sum_{i=1}^{N}\left|\gamma_{N,t}^{(i)}-\gamma_{N,t}^{(i-1)}\right|-\sum_{i=1}^{N-1}\operatorname{sgn}\left(\gamma_{N,t}^{(i+1)}-\gamma_{N,t}^{(i)}\right)\left(\gamma_{N,t}^{(i)}-\gamma_{N,t}^{(i-1)}\right)\\
&\geq\sum_{i=1}^{N-1}\left|\gamma_{N,t}^{(i)}-\gamma_{N,t}^{(i-1)}\right|-\sum_{i=1}^{N-1}\operatorname{sgn}\left(\gamma_{N,t}^{(i+1)}-\gamma_{N,t}^{(i)}\right)\left(\gamma_{N,t}^{(i)}-\gamma_{N,t}^{(i-1)}\right)\\
&=\sum_{i=1}^{N-1}\left(\left|\gamma_{N,t}^{(i)}-\gamma_{N,t}^{(i-1)}\right|-\operatorname{sgn}\left(\gamma_{N,t}^{(i+1)}-\gamma_{N,t}^{(i)}\right)\left(\gamma_{N,t}^{(i)}-\gamma_{N,t}^{(i-1)}\right)\right)\\
&\geq 0.
\end{aligned}
\tag{104}
$$

Consequently, we obtain

$$
\begin{aligned}
\frac{\mathrm{d}}{\mathrm{d}t}\sum_{i=1}^{N}\left|\gamma_{N,t}^{(i)}-\gamma_{N,t}^{(i-1)}\right|&=\frac{\mathrm{d}}{\mathrm{d}t}\left(\sum_{i=2}^{N}\left|\gamma_{N,t}^{(i)}-\gamma_{N,t}^{(i-1)}\right|+\left|\gamma_{N,t}^{(1)}\right|\right)\\
&\geq\frac{u}{h}\left(\sum_{i=2}^{N}\operatorname{sgn}\left(\gamma_{N,t}^{(i)}-\gamma_{N,t}^{(i-1)}\right)\left(\gamma_{N,t}^{(i)}-2\gamma_{N,t}^{(i-1)}+\gamma_{N,t}^{(i-2)}\right)-\gamma_{N,t}^{(1)}\right)-\sum_{i=2}^{N}C_{r,\sigma}h\|\lambda\|_{\infty}\\
&\geq -C_{r,\sigma}L\|\lambda\|_{\infty}.
\end{aligned}
\tag{105}
$$

Hence, we arrive at the desired bound:

$$
\begin{aligned}
\mathrm{TV}\left(\gamma_{N,t}\right)&=\sum_{i=2}^{N}\left|\gamma_{N,t}^{(i)}-\gamma_{N,t}^{(i-1)}\right|+\left|\gamma_{N,t}^{(1)}\right|\\
&\leq\sum_{i=2}^{N}\left|\gamma_{N,T}^{(i)}-\gamma_{N,T}^{(i-1)}\right|+\left|\gamma_{N,T}^{(1)}\right|+C_{r,\sigma}\|\lambda\|_{\infty}L(T-t)\\
&=h\sum_{i=2}^{N}\left|\frac{\lambda^{(i)}-\lambda^{(i-1)}}{h}\right|+\left|\lambda_{N,T}^{(1)}\right|+C_{r,\sigma}\|\lambda\|_{\infty}L(T-t)\\
&\leq L\|\lambda'\|_{\infty}+\|\lambda\|_{\infty}+C_{r,\sigma}\|\lambda\|_{\infty}LT.
\end{aligned}
\tag{106}
$$

This shows, for any $t_1,t_2\in[0,T]$, that

$$
h\sum_{i=1}^{N}\left|\frac{\mathrm{d}\gamma_{N,t}^{(i)}}{\mathrm{d}t}\right|\leq C''\left(\mathrm{TV}\left(\gamma_{N,t}\right)+1\right)\leq C''\left(L\|\lambda'\|_{\infty}+\|\lambda\|_{\infty}+C_{r,\sigma}\|\lambda\|_{\infty}LT+1\right)(=C''')
\tag{107}
$$

with a constant $C''>0$ independent of $N$. This implies

$$
h\sum_{i=1}^{N}\left|\gamma_{N,t_1}^{(i)}-\gamma_{N,t_2}^{(i)}\right|\leq C'''\left|t_1-t_2\right|,\quad 0\leq t_1,t_2\leq T.
\tag{108}
$$

**Step 2: Convergence to $\gamma$**

These observations (consistency, global boundedness, TV is bounded, and comparison) imply the following convergence result (e.g., Theorem 15.2 in LeVeque [87]):

$$
\lim_{N\to+\infty}\left|\gamma_{N,t}^{(i)}-\gamma_t\left(x_i\right)\right|=0\quad\text{in}\quad L^1\left([0,T]\times D\right)
\tag{109}
$$

with $\gamma$, satisfying the uniform bound (87), such that

$$0=\int_0^L \phi(x)\left(\frac{\partial \gamma_t(x)}{\partial t}-r_t(x)\gamma_t(x)+\frac{1}{2}\left(\sigma_t(x)\gamma_t(x)\right)^2\right)\mathrm{d}x+\int_0^L u\frac{\mathrm{d}\phi(x)}{\mathrm{d}x}\gamma_t(x)\mathrm{d}x-u\phi(L)\gamma_t(L) \quad (110)$$

for all $\phi\in C^\infty(\bar{D})$, which comes from the limit $N\to+\infty$ of the equation

$$\sum_{i=1}^N\left(\left(\frac{\mathrm{d}\gamma_{N,t}^{(i)}}{\mathrm{d}t}-r_t^{(i)}\gamma_{N,t}^{(i)}+\frac{1}{2}\left(\gamma_{N,t}^{(i)}\right)^2\left(\sigma_t^{(i)}\right)^2\right)\phi(x_i)+u\sum_{i=1}^N\gamma_{N,t}^{(i)}\frac{\phi(x_{i+1})-\phi(x_i)}{h}\right)-u\gamma_{N,t}^{(N)}\phi(x_N)=0\ , \quad (111)$$

where

$$-\sum_{i=1}^N\phi(x_i)\frac{\gamma_{N,t}^{(i)}-\gamma_{N,t}^{(i-1)}}{h}=\sum_{i=1}^N\gamma_{N,t}^{(i)}\frac{\phi(x_{i+1})-\phi(x_i)}{h}-\gamma_{N,t}^{(N)}\phi(x_N)\,. \quad (112)$$

Moreover, the obtained $\gamma$ satisfies the uniform bound $-\lambda\le\gamma\le 0$. This $\gamma$ is an entropy solution due to consistency and monotonicity (Theorem 15.7 in LeVeque [87]). The uniqueness of $\gamma$ then follows from the linearity of the advection term and local Lipschitz continuity of the source term $r\gamma-\sigma^2\gamma^2/2$ along with the uniform bound $-\lambda\le\gamma\le 0$; we can use the uniqueness result of entropy solutions for conservation laws with boundary data and source based on the Kružkov's framework because of $r0-\sigma^2 0^2/2=0$ and $r(-\lambda)-\sigma^2(-\lambda)^2/2\le 0$ (e.g., Assumption 1 and Theorem 18 in Martin [86]).

**Step 3: Remaining coefficients**

The convergence from $\alpha_{N,\cdot}\to\alpha_\cdot$ follows since $\beta_{N,\cdot}\to\beta_\cdot$ and $\beta\in C^1(D)$. The fact that $\beta\in C^1(D)$ follows because the following function of $t$,

$$\int_0^L \omega_{t,t}(x)\mathrm{d}x=\int_0^L \gamma_{t+x/v}(x)g_{t+x/v}(x)\mathbb{I}(t<T-x/v)\mathrm{d}x=\int_0^{\min\{L,v(T-t)\}}\gamma_{t+x/v}(x)g_{t+x/v}(x)\mathrm{d}x \quad (113)$$

is continuous for $0\le t\le T$ due to the integrability of $\gamma$.

□

***Proof of Proposition 5***

We follow the strategy analogous to that employed in Proof of Theorem 3.1 in Bansaye and Méléard [89]. For $\mathbb{Y}_{N,t}(\phi)$, the relation (29) holds true with $x_0=0$ instead of $\phi(x_0)=0$, because the additional term $\phi(0)\mathbb{F}_{N,t}$ compensates the term for $i=1$.

We evaluate each term of (29). We have

$$\begin{aligned}\mathbb{E}\left[\left|M_{N,S}(\phi)-M_{N,S'}(\phi)\right|^2\right]&\le\mathbb{E}\left[\int_S^{\min\{S+\eta,T\}}h\sum_{i=1}^N Y_s^{(i)}\left(\phi(x_i)\right)^2\left(\sigma_s^i\right)^2\mathrm{d}s\right]\\&\le\left(\|\sigma\|_\infty\|\phi\|_\infty\right)^2\mathbb{E}\left[\sup_{0\le t\le T}h\sum_{i=1}^N Y_t^{(i)}\right]\eta \quad (114)\\&\underset{\eta\to 0}{\to}0,\end{aligned}$$

where the limit is uniform. Here, we used $\mathbb{E}\left[\sup_{0\le t\le T} h\sum_{i=1}^{N} Y_t^{(i)}\right] \le \mathbb{E}\left[\left(\sup_{0\le t\le T} h\sum_{i=1}^{N} Y_t^{(i)}\right)^2\right]^{\frac{1}{2}} \le \sqrt{C_T}$ coming from (148).

For the other terms in (29), we have

$$\mathbb{E}\left[\left|\int_S^{S'}\left(-uh\sum_{i=1}^{N} Y_s^{(i)} \frac{\phi(x_i)-\phi(x_{i-1})}{h}\right)\mathrm{d}s\right|\right] \le \|\phi'\|_\infty u\sqrt{C_T}\,\eta \underset{\eta\to 0}{\to} 0, \tag{115}$$

$$\mathbb{E}\left[\left|\int_S^{S'} h\sum_{i=1}^{N} g_s^{(i)} Z_{s-x_i/v}\phi(x_i)\mathrm{d}s\right|\right] \le L\mathbb{E}\left[\sup_{t\in\mathbb{R}} Z_t\right]\|g\|_\infty \|\phi\|_\infty\, \eta \underset{\eta\to 0}{\to} 0\,, \tag{116}$$

$$\mathbb{E}\left[\left|\int_S^{S'} h\sum_{i=1}^{N} r_s^{(i)} Y_s^{(i)}\phi(x_i)\mathrm{d}s\right|\right] \le \sqrt{C_T}\,\|r\|_\infty \|\phi\|_\infty\, \eta \underset{\eta\to 0}{\to} 0\,, \tag{117}$$

where the limits are uniform ((116) is due to the estimate (137)). The observations (114) through (117) along with (29) show that the sequence $\mathbb{Y}_{N,\cdot}(\phi)$ ($N = 1,2,3,\ldots$) satisfies the Aldous criterion (33) (e.g., Eq. (3.10) in Rudnicki and Wieczorek [64]; Chapter 3.1, p. 21 in Bansaye and Méléard [89]).

□

***Proof of Proposition 6***

**Step 1: Several estimates**

We choose a test function $\phi = \phi_{l,k} \in C^\infty(\bar{D})$ ($k\in\mathbb{N}$) such that $\phi_{l,k}(0)=0$, which is nonnegative, nondecreasing, bounded between 0 and 1, and approaches the indicator function $\mathbb{I}(D_l)(\cdot)$ for the set $D_l$ as $k\to+\infty$ from above (i.e., $\phi_{k,l}\ge \mathbb{I}(D_l)$). Then, we have

$$\begin{aligned}
&\mathbb{E}\left[\sup_{t\le T} h\sum_{i=1}^{N} Y_t^{(i)}\phi_{l,k}(x_i)\right] \\
&\le \mathbb{E}\left[\sup_{t\le T}\int_0^t\left(-uh\sum_{i=1}^{N} Y_s^{(i)}\frac{\phi_{l,k}(x_i)-\phi_{l,k}(x_{i-1})}{h}\right)\mathrm{d}s\right] + \|g\|_\infty\, \mathbb{E}\left[\sup_{t\le T}\int_0^t h\sum_{i=1}^{N} Z_{s-x_i/v}\phi_{l,k}(x_i)\mathrm{d}s\right] \\
&+\mathbb{E}\left[\sup_{t\le T}\left|M_{N,t}(\phi_{l,k})\right|\right].
\end{aligned} \tag{118}$$

By the nondecreasing nature of $\phi_{l,k}$, we have

$$\frac{\phi_{l,k}(x_i)-\phi_{l,k}(x_{i-1})}{h} \ge 0 \quad \text{for all } i\in I \text{ and } k\in\mathbb{N}\,, \tag{119}$$

implying that

$$\mathbb{E}\left[\sup_{t\le T} h\sum_{i=1}^{N} Y_t^{(i)}\phi_{l,k}(x_i)\right] \le \|g\|_\infty\, \mathbb{E}\left[\sup_{t\le T}\int_0^t h\sum_{i=1}^{N} Z_{s-x_i/v}\phi_{l,k}(x_i)\mathrm{d}s\right] + \mathbb{E}\left[\sup_{t\le T}\left|\int_0^t M_{N,t}(\phi_{l,k})\mathrm{d}s\right|\right]. \tag{120}$$

Applying Burkholder's inequality (Theorem 7.3 in Mao [88]) along with Jensen's equality to the last term yields

$$
\begin{aligned}
\left(\mathbb{E}\left[\sup_{t\le T}\left|M_{N,t}\left(\phi_{l,k}\right)\right|\right]\right)^2 &\le \mathbb{E}\left[\sup_{t\le T}\left|M_{N,t}\left(\phi_{l,k}\right)\right|^2\right] \\
&\le C_2\mathbb{E}\left[\left\langle M_{N,T}\left(\phi_{l,k}\right)\right\rangle\right] \\
&\le C_2\left(\left\|\sigma\right\|_\infty\right)^2\mathbb{E}\left[\int_0^T h\sum_{i=1}^N Y_s^{(i)}\left(\phi_{l,k}\left(x_i\right)\right)^2 \mathrm{d}s\right] \\
&= C_2\left(\left\|\sigma\right\|_\infty\right)^2\int_0^T h\sum_{i=1}^N \mathbb{E}\left[Y_s^{(i)}\right]\left(\phi_{l,k}\left(x_i\right)\right)^2 \mathrm{d}s \\
&\le C_2C_YT\left(\left\|\sigma\right\|_\infty\right)^2 h\sum_{i=1}^N\left(\phi_{l,k}\left(x_i\right)\right)^2
\end{aligned}
\tag{121}
$$

with a global constant $C_2 > 0$ and a constant $C_Y > 0$; the existence of the latter constant is separately proven later in the last step. For the first term on the right-hand side of (120), we have

$$
\mathbb{E}\left[\sup_{t\le T}\int_0^t h\sum_{i=1}^N Z_{s-x_i/v}\phi_{l,k}\left(x_i\right)\mathrm{d}s\right] \le \mathbb{E}\left[h\sum_{i=1}^N\phi_{l,k}\left(x_i\right)\left(\int_0^T\sup_{t\in\mathbb{R}}Z_t\mathrm{d}s\right)\right] = T\mathbb{E}\left[\sup_{t\in\mathbb{R}}Z_t\right]h\sum_{i=1}^N\phi_{l,k}\left(x_i\right). \tag{122}
$$

**Step 2: Decompose $Z$**

By (46) and Jensen's inequality, we have

$$
\begin{aligned}
\mathbb{E}\left[\sup_{t\in\mathbb{R}}Z_t\right] &\le \mathbb{E}\left[\sup_{t\in\mathbb{R}}A_{Z,t}\right] + \mathbb{E}\left[\sup_{t\in\mathbb{R}}\left|M_{Z,t}\right|\right] \\
&= \sup_{0\le t<\tau}\int_0^\tau a_s\left(\frac{\tau-t}{\tau-s}\right)^b\mathrm{d}s + \mathbb{E}\left[\sup_{0\le t<\tau}\left|M_{Z,t}\right|\right] \\
&\le \sup_{0\le t<\tau}\int_0^\tau a_s\left(\frac{\tau-t}{\tau-s}\right)^b\mathrm{d}s + \left(\mathbb{E}\left[\sup_{0\le t<\tau}\left|M_{Z,t}\right|^2\right]\right)^{\frac{1}{2}}.
\end{aligned}
\tag{123}
$$

The first term in (123) is $\sup_{0\le t\le\tau}\mathbb{E}\left[Z_t\right]\left(\le\bar{Z}\right)$. For later use, for any $\varepsilon\in\left(0,\tau\right)$, we set the local martingale

$$
m_{Z,t} = \int_0^t c_s\left(\frac{1}{\tau-s}\right)^b\sqrt{\frac{b}{\tau-s}Z_s}\mathrm{d}B_s = b^{\frac{1}{2}}\int_0^t c_s\left(\tau-s\right)^{-b-\frac{1}{2}}\sqrt{Z_s}\mathrm{d}B_s\ ,\ \ 0<t<\tau \tag{124}
$$

and obtain

$$
M_{Z,t} = \left(\tau-t\right)^b m_{Z,t} = \left(\tau-t\right)^b m_{Z,\tau-\varepsilon} + \left(\tau-t\right)^b\left(m_{Z,t}-m_{Z,\tau-\varepsilon}\right). \tag{125}
$$

Clearly, we have

$$
\sup_{0\le t<\tau}\left|M_{Z,t}\right|^2 \le \sup_{0\le t\le\tau-\varepsilon}\left|M_{Z,t}\right|^2 + \sup_{\tau-\varepsilon\le t<\tau}\left|M_{Z,t}\right|^2. \tag{126}
$$

We also note the estimate that will be used later (we temporarily assume $b\ne1$, but the case $b=1$ follows in an analogous way where $\min\left\{b,1\right\}$ in (127) needs to be replaced by $1-\vartheta$ with a sufficiently small $\vartheta\in\left(0,1\right)$) with some constant $C_{a,b}>0$ depending only on $a,b$:

$$
\mathbb{E}\left[Z_t\right] = \int_0^t a_s\left(\frac{\tau-t}{\tau-s}\right)^b\mathrm{d}s \le \left\|a\right\|_\infty\int_0^t\left(\frac{\tau-t}{\tau-s}\right)^b\mathrm{d}s \le C_{a,b}\left(\tau-t\right)^{\min\left\{b,1\right\}}\ ,\ \ 0<t<\tau. \tag{127}
$$

**Step 3: Estimate fluctuations in $Z$**

For the first term in (126), we have

$$\mathbb{E}\left[\sup_{0\le t\le \tau-\varepsilon}\left|(\tau-t)^b m_{Z,t}\right|^2\right]\le \tau^{2b}\mathbb{E}\left[\sup_{0\le t\le \tau-\varepsilon}\left|b^{\frac{1}{2}}\int_0^t c_s(\tau-s)^{-b-\frac{1}{2}}\sqrt{Z_s}\mathrm{d}B_s\right|^2\right]$$
$$\le b\tau^{2b}C_2\int_0^{\tau-\varepsilon}(c_s)^2(\tau-s)^{-2b-1}\mathbb{E}[Z_s]\mathrm{d}s \tag{128}$$
$$\le b\tau^{2b}\bar{Z}\left(\|c\|_\infty\right)^2 C_2\int_0^{\tau-\varepsilon}(\tau-s)^{-2b-1}\mathrm{d}s$$
$$<+\infty.$$

For the second term of (126), we use the decomposition (125). First, we have

$$\mathbb{E}\left[\sup_{\tau-\varepsilon\le t<\tau}\left|(\tau-t)^b m_{Z,\tau-\varepsilon}\right|^2\right]\le \varepsilon^{2b}\mathbb{E}\left[\left|m_{Z,\tau-\varepsilon}\right|^2\right]<+\infty. \tag{129}$$

Second, for $\psi_t = m_{Z,t}-m_{Z,\tau-\varepsilon}$, we have the integration-by-parts formula ($\mathrm{d}\psi_s\mathrm{d}s=0$):

$$\begin{aligned}(\tau-t)^b\psi_t &= \int_{\tau-\varepsilon}^t \mathrm{d}\left((\tau-s)^b\psi_s\right)\\ &= \int_{\tau-\varepsilon}^t(\tau-s)^b\mathrm{d}\psi_s+\int_{\tau-\varepsilon}^t\psi_s\mathrm{d}\left((\tau-s)^b\right)\\ &= \underbrace{\int_{\tau-\varepsilon}^t(\tau-s)^b\mathrm{d}\psi_s}_{E_{1,t}}\underbrace{-b\int_{\tau-\varepsilon}^t(\tau-s)^{b-1}\psi_s\mathrm{d}s}_{E_{2,t}}.\end{aligned} \tag{130}$$

Then, we obtain

$$\mathbb{E}\left[\sup_{\tau-\varepsilon\le t<\tau}\left|(\tau-t)^b\psi_t\right|^2\right]\le 2\mathbb{E}\left[\sup_{\tau-\varepsilon\le t<\tau}\left(E_{1,t}\right)^2\right]+2\mathbb{E}\left[\sup_{\tau-\varepsilon\le t<\tau}\left(E_{2,t}\right)^2\right]. \tag{131}$$

We evaluate each term of (131). For the first term, by Doob's maximal inequality (Theorem 3.8(iv) in Karatzas and Shreve [85]) and the estimate (127), we obtain

$$\begin{aligned}\mathbb{E}\left[\sup_{\tau-\varepsilon\le t<\tau}\left(E_{1,t}\right)^2\right]&\le 4\mathbb{E}\left[\left(\int_{\tau-\varepsilon}^\tau(\tau-s)^b\mathrm{d}\psi_s\right)^2\right]\\ &=4\mathbb{E}\left[\left(\int_{\tau-\varepsilon}^\tau(\tau-s)^b\left(b^{\frac{1}{2}}c_s(\tau-s)^{-b-\frac{1}{2}}\sqrt{Z_s}\mathrm{d}B_s\right)\right)^2\right]\\ &\le 4b\left(\|c\|_\infty\right)^2\mathbb{E}\left[\left(\int_{\tau-\varepsilon}^\tau(\tau-s)^{-\frac{1}{2}}\sqrt{Z_s}\mathrm{d}B_s\right)^2\right]\\ &=4b\left(\|c\|_\infty\right)^2\int_{\tau-\varepsilon}^\tau(\tau-s)^{-1}\mathbb{E}[Z_s]\mathrm{d}s\\ &\le 4C_{a,b}b\left(\|c\|_\infty\right)^2\int_{\tau-\varepsilon}^\tau(\tau-s)^{\min\{b,1\}-1}\mathrm{d}s\\ &<+\infty.\end{aligned} \tag{132}$$

For the second term, we note that

$$\sup_{\tau-\varepsilon\le t<\tau}\left|E_{2,t}\right|=b\sup_{\tau-\varepsilon\le t<\tau}\left|\int_{\tau-\varepsilon}^t(\tau-s)^{b-1}\psi_s\mathrm{d}s\right|\le b\sup_{\tau-\varepsilon\le t<\tau}\int_{\tau-\varepsilon}^t(\tau-s)^{b-1}\left|\psi_s\right|\mathrm{d}s\le b\int_{\tau-\varepsilon}^\tau(\tau-s)^{b-1}\left|\psi_s\right|\mathrm{d}s. \tag{133}$$

Then, because the integral of a norm is not smaller than the norm of the integral, we can proceed as follows:

$$
\begin{aligned}
\left(\mathbb{E}\left[\sup_{\tau-\varepsilon\le t<\tau}\left(E_{2,t}\right)^2\right]\right)^{\frac{1}{2}} &\le b\left(\mathbb{E}\left[\left(\int_{\tau-\varepsilon}^{\tau}(\tau-s)^{b-1}|\psi_s|\mathrm{d}s\right)^2\right]\right)^{\frac{1}{2}} \\
&\le b\int_{\tau-\varepsilon}^{\tau}\left(\mathbb{E}\left[\left((\tau-s)^{b-1}|\psi_s|\right)^2\right]\right)^{\frac{1}{2}}\mathrm{d}s \\
&= b\int_{\tau-\varepsilon}^{\tau}(\tau-s)^{b-1}\left(\mathbb{E}\left[(\psi_s)^2\right]\right)^{\frac{1}{2}}\mathrm{d}s \\
&= b\int_{\tau-\varepsilon}^{\tau}(\tau-s)^{b-1}\left(\mathbb{E}\left[\left(b^{\frac{1}{2}}\int_0^s c_u(\tau-u)^{-b-\frac{1}{2}}\sqrt{Z_u}\,\mathrm{d}B_u\right)^2\right]\right)^{\frac{1}{2}}\mathrm{d}s \\
&= b^{\frac{3}{2}}\int_{\tau-\varepsilon}^{\tau}(\tau-s)^{b-1}\left(\int_0^s(c_u)^2(\tau-u)^{-2b-1}\,\mathbb{E}[Z_u]\mathrm{d}u\right)^{\frac{1}{2}}\mathrm{d}s \\
&\le b^{\frac{3}{2}}\|c\|_\infty\left(C_{a,b}\right)^{\frac{1}{2}}\int_{\tau-\varepsilon}^{\tau}(\tau-s)^{b-1}\left(\int_0^s(\tau-u)^{\min\{b,1\}-2b-1}\mathrm{d}u\right)^{\frac{1}{2}}\mathrm{d}s.
\end{aligned}
\tag{134}
$$

Here, we used (127) to obtain the last line in (134). Because $b>0$, we have

$$
\begin{aligned}
\int_{\tau-\varepsilon}^{\tau}(\tau-s)^{b-1}\left(\int_0^s(\tau-u)^{\min\{b,1\}-2b-1}\mathrm{d}u\right)^{\frac{1}{2}}\mathrm{d}s &= \int_{\tau-\varepsilon}^{\tau}(\tau-s)^{b-1}\left(O\left((\tau-s)^{\min\{b,1\}-2b}\right)\right)^{\frac{1}{2}}\mathrm{d}s \\
&= \int_{\tau-\varepsilon}^{\tau}O\left((\tau-s)^{\frac{1}{2}\min\{b,1\}-1}\right)\mathrm{d}s \\
&< +\infty.
\end{aligned}
\tag{135}
$$

Consequently, we obtain

$$
\mathbb{E}\left[\sup_{\tau-\varepsilon\le t<\tau}\left|(\tau-t)^b\left(m_{Z,t}-m_{Z,\tau-\varepsilon}\right)\right|^2\right]<+\infty \tag{136}
$$

and hence, by (123) and (126),

$$
\mathbb{E}\left[\sup_{t\in\mathbb{R}}Z_t\right]<+\infty. \tag{137}
$$

**Step 4: Desired estimate**

In summary, we obtain

$$
\mathbb{E}\left[\sup_{t\le T}h\sum_{i=1}^{N}Y_t^{(i)}\phi_{l,k}(x_i)\right]\le\|g\|_\infty C_Z h\sum_{i=1}^{N}\phi_{l,k}(x_i)+\sqrt{C_2C_Y\left(\|\sigma\|_\infty\right)^2h\sum_{i=1}^{N}\left(\phi_{l,k}(x_i)\right)^2}\left(=F\left(h\sum_{i=1}^{N}\phi_{l,k}(x_i)\right)\right) \tag{138}
$$

with a constant $C_Z>0$ depending on $Z,\varepsilon$ and a nonnegative and continuous function $F$ with $F(0)=0$. By the choice of $\phi_{l,k}$, we have

$$
\mathbb{E}\left[\sup_{t\le T}\mu_{N,t}(D_l)\right]\le\mathbb{E}\left[\sup_{t\le T}h\sum_{i=1}^{N}Y_t^{(i)}\phi_{l,k}(x_i)\right]\le F\left(h\sum_{i=1}^{N}\phi_{l,k}(x_i)\right) \tag{139}
$$

Letting $k\to+\infty$ yields

$$
\mathbb{E}\left[\sup_{t\le T}\mu_{N,t}(D_l)\right]\le F\left(h\sum_{i=1}^{N}\mathbb{I}(D_l)(x_i)\right). \tag{140}
$$

Fix arbitrary $\eta>0$. By Markov's inequality (Theorem 5.11 in Klenke (2020)[84]), we have

$$\mathbb{P}\left(\sup_{t\le T}\mu_{N,t}\left(D_l\right)>\eta\right)\le\frac{1}{\eta}\mathbb{E}\left[\sup_{t\le T}\mu_{N,t}\left(D_l\right)\right]\le\frac{1}{\eta}F\left(h\sum_{i=1}^{N}\mathbb{I}\left(D_l\right)\left(x_i\right)\right). \tag{141}$$

We therefore obtain

$$\sup_N\mathbb{P}\left(\sup_{t\le T}\mu_{N,t}\left(D_l\right)>\eta\right)\le\frac{1}{\eta}\sup_N F\left(h\sum_{i=1}^{N}\mathbb{I}\left(D_l\right)\left(x_i\right)\right). \tag{142}$$

For a large $l>0$, the right-hand side of (142) is scaled as $\eta^{-1}l^{-1/2}$; hence, we can always choose a sufficiently large $l>0$ that possibly depends on $\eta$ such that (34) is satisfied.

**Step 5: Existence of $C_Y$**

We show the postponed proof that $\sup_{t\le T,i\in I}\mathbb{E}\left[Y_t^{(i)}\right]\le C_Y\left(=ZL\|g\|_\infty/u\right)$. We have

$$\frac{\mathrm{d}}{\mathrm{d}t}\mathbb{E}\left[Y_t^{(N)}\right]\le-\frac{u}{h}\mathbb{E}\left[Y_t^{(N)}\right]+g_t^{(N)}\mathbb{E}\left[Z_{t-\tau_i}\right]\le-\frac{u}{h}\mathbb{E}\left[Y_t^{(N)}\right]+\bar{Z}\|g\|_\infty, \tag{143}$$

and hence, by using Gronwall's inequality,

$$\mathbb{E}\left[Y_t^{(N)}\right]\le\bar{Z}\|g\|_\infty\frac{h}{u}\left(1-e^{-ut/h}\right)\le\frac{\bar{Z}\|g\|_\infty}{u}h. \tag{144}$$

We also have

$$\frac{\mathrm{d}}{\mathrm{d}t}\mathbb{E}\left[Y_t^{(N-1)}\right]\le-\frac{u}{h}\mathbb{E}\left[Y_t^{(N-1)}\right]+\frac{u}{h}\mathbb{E}\left[Y_t^{(N)}\right]+\bar{Z}\|g\|_\infty,\ t>0 \tag{145}$$

and hence

$$\begin{aligned}\mathbb{E}\left[Y_t^{(N-1)}\right]&\le\int_0^t e^{-u(t-s)/h}\left(\bar{Z}\|g\|_\infty\left(1-e^{-us/h}\right)+\bar{Z}\|g\|_\infty\right)\mathrm{d}s\\&\le2\bar{Z}\|g\|_\infty\int_0^t e^{-u(t-s)/h}\mathrm{d}s\\&=2Z\|g\|_\infty\frac{h}{u}\left(1-e^{-ut/h}\right)\\&\le\frac{\bar{Z}\|g\|_\infty}{u}(2h).\end{aligned} \tag{146}$$

By induction, iterating this procedure to $i=1$ yields the desired result:

$$\mathbb{E}\left[Y_t^{(i)}\right]\le\frac{\bar{Z}}{u}\|g\|_\infty\left(\left(N-i+1\right)h\right)\le\frac{\bar{Z}L\|g\|_\infty}{u}<+\infty,\ i\in I. \tag{147}$$

□

***Proof of Proposition 7***

This is a consequence of Conditions (i) and (ii) in Theorem II.4.1 of Perkins [65] applied to the process $\mu_{N,t}\left(\mathrm{d}x\right)+\mathbb{F}_{N,\cdot}\delta\left(x\right)$ regarded as a measure on $\left[0,L\right)$, which coincides with $\mu_{N,t}$ on $D$; the former condition was verified in **Proposition 6** because we could find one suitable set (i.e., $D_l$) and the latter by the tightness shown in **Proposition 5** (because $\mathbb{Y}_{N,\cdot}\left(\cdot\right)$ satisfies the Aldous criterion (33) and is continuous in $t$ due to the uniform $\eta$-limits shown in **Proof of Proposition 5**). The existence of its limit under $N \to +\infty$ in law, up to subsequence, would follow Prokhorov's theorem (Theorem 13.29 in Klenke [84]). Applying dominated convergence to each term in (29) along with their continuous dependence on such a subsequence $Y_t^{(i)}$ can yield that there exists a weak solution $\mu$ to the SPDE model, and further $\langle M_{N,t}(\cdot)\rangle \to \langle M_t(\cdot)\rangle$ in probability. These observations combined with the tightness argument analogous to that in Proposition II.4.2 (p.31) of Perkins [65] imply the convergence in probability, of each term in (29) to the corresponding term of (9) as well as the convergence of $\mathbb{F}_{N,\cdot}$ due to the second estimate in (148) below.

To apply these results (Prokhorov's theorem and tightness), we must show

$$\mathbb{E}\left[\sup_{t\le T}\left(h\sum_{i=1}^{N} Y_t^{(i)}\right)^2\right],\ \mathbb{E}\left[\sup_{t\le T}\left(\int_0^t uY_s^{(1)}\mathrm{d}s\right)^2\right] \le C_T \tag{148}$$

for a constant $C_T > 0$ irrespective of $N \in \mathbb{N}$. First, we have

$$h\sum_{i=1}^{N} Y_t^{(i)} + \int_0^t uY_s^{(1)}\mathrm{d}s \le \int_0^t h\sum_{i=1}^{N} g_s^{(i)} Z_{s-x_i/v}\mathrm{d}s + M_{N,t}(1) \le \int_0^t h\sum_{i=1}^{N} g_s^{(i)} Z_{s-x_i/v}\mathrm{d}s + M_{N,t}(1). \tag{149}$$

From (149), we obtain

$$\begin{aligned}\sup_{t\le T}\left(h\sum_{i=1}^{N} Y_t^{(i)}\right)^2,\ \sup_{t\le T}\left(\int_0^t uY_s^{(1)}\mathrm{d}s\right)^2 &\le 2\sup_{t\le T}\left(\left(\int_0^t h\sum_{i=1}^{N} g_s^{(i)} Z_{s-x_i/v}\mathrm{d}s\right)^2 + \left(M_{N,t}(1)\right)^2\right)\\ &\le 2\left(\|g\|_\infty\right)^2 T\sup_{t\le T}\int_0^t\left(h\sum_{i=1}^{N} Z_{s-x_i/v}\right)^2\mathrm{d}s + 2\sup_{t\le T}\left(M_{N,t}(1)\right)^2\\ &\le 2\left(\|g\|_\infty\right)^2 T\int_0^T\left(h\sum_{i=1}^{N} Z_{s-x_i/v}\right)^2\mathrm{d}s + 2\sup_{t\le T}\left(M_{N,t}(1)\right)^2.\end{aligned} \tag{150}$$

For the second term in (150), by the estimate analogous to (121) (we simply replace $\phi_{k,l}$ by $1$), we obtain $\sup_{N\in\mathbb{N}}\mathbb{E}\left[\sup_{t\le T}\left(M_{N,t}(1)\right)^2\right] < +\infty$. For the first term, we have

$$\mathbb{E}\left[\int_0^T\left(h\sum_{i=1}^{N} Z_{s-x_i/v}\right)^2\mathrm{d}s\right] \le \mathbb{E}\left[\int_0^T\sup_{s\in\mathbb{R}}\left(h\sum_{i=1}^{N} Z_s\right)^2\mathrm{d}s\right] = TL^2\mathbb{E}\left[\sup_{t\in\mathbb{R}}\left(Z_t\right)^2\right] \tag{151}$$

and

$$\mathbb{E}\left[\sup_{t\in\mathbb{R}}\left(Z_t\right)^2\right] \le 2\left(\sup_{0<t<\tau}\left|A_{Z,t}\right|^2 + \mathbb{E}\left[\sup_{0<t<\tau}\left|M_{Z,t}\right|^2\right]\right) \tag{152}$$

and then use the boundedness results of the right-hand side of (152) as shown in **Proof of Proposition 6**, proving (148). Finally, applying Theorem 1.19 in Li [66] to our SPDE model along with **Proposition 4** yields the uniqueness in law.

□

### A.2 A lemma

The following lemma proves the convergence argument of $Z$ at $\tau$ in **Step 1** of **Proof of Proposition 1**. It also generalizes (and corrects the "martingale argument" in) Proof of Proposition 1 (below Eq. (37)) in Yoshioka [46]. We work on a complete filtered probability space $\left(\Omega, \mathcal{F}, (\mathcal{F}_t)_{t\geq 0}, \mathbb{P}\right)$.

***Lemma A.1***

*Fix* $\tau > 0$. *Consider the process*

$$z_t = \int_0^t c_s \left(\frac{\tau - t}{\tau - s}\right)^b \sqrt{\frac{1}{\tau - s}\xi_s}\,\mathrm{d}B_s,\quad 0 \leq t \leq \tau, \tag{153}$$

*where* $b > 0$, $(B_t)_{t\geq 0}$ *is a 1-D standard Brownian motion,* $(c_t)_{0\leq t\leq \tau}$ *is a bounded and continuous function, and* $(\xi_t)_{0\leq t<\tau}$ *is a continuous and nonnegative* $\mathcal{F}_t$*-predictable process such that*

$$\mathbb{E}[\xi_t] \leq C(\tau - t)^{\min\{b,1\}},\quad 0 < t < \tau \tag{154}$$

*with a constant* $C > 0$. *Then,* $z_t \to 0$ *as* $t \nearrow \tau$ *with probability 1.*

***Proof of Lemma A.1***

For $0 \leq t < \tau$, we write

$$z_t = (\tau - t)^b \int_0^t c_s \left(\frac{1}{\tau - s}\right)^b \sqrt{\frac{1}{\tau - s}\xi_s}\,\mathrm{d}B_s = (\tau - t)^b y_t \tag{155}$$

with $(y)_{0\leq t<\tau}$ being a local martingale. By (154), we have

$$\mathbb{E}\left[(y_t)^2\right] = \int_0^t (c_s)^2 \frac{1}{(\tau - s)^{2b+1}}\mathbb{E}[\xi_s]\mathrm{d}s \leq C\left(\|c\|_\infty\right)^2 \int_0^t (\tau - s)^{\min\{b,1\}-2b-1}\mathrm{d}s. \tag{156}$$

Next, we fix any $\varepsilon > 0$, and then take $t_n = \tau(1 - 2^{-n})$ and $A_n = \left\{\sup_{t\in[t_n, t_{n+1}]} |z_t| \geq \varepsilon\right\}$ ($n = 1, 2, 3, \ldots$). Because $\tau - t_n = 2(\tau - t_{n+1})$, we have

$$\mathbb{P}(A_n) = \mathbb{P}\left(\sup_{t\in[t_n,t_{n+1}]} (\tau - t)^b |y_t| \geq \varepsilon\right) \leq \mathbb{P}\left(\sup_{t\in[t_n,t_{n+1}]} (\tau - t)^b \sup_{t\leq t_{n+1}} |y_t| \geq \varepsilon\right) = \mathbb{P}\left(\sup_{t\leq t_{n+1}} |y_t| \geq \varepsilon 2^{-b} (\tau - t_{n+1})^{-b}\right). \tag{157}$$

Then, Markov's inequality (Theorem 5.11 in Klenke [84]) along with Doob's maximal inequality (Theorem 3.8(iv) in Karatzas and Shreve [85]) shows

$$\mathbb{P}(A_n) \leq \frac{\mathbb{E}\left[\sup_{t\leq t_{n+1}} |y_t|^2\right]}{\left(\varepsilon 2^{-b}(\tau - t_{n+1})^{-b}\right)^2} \leq \frac{4\mathbb{E}\left[\left(y_{t_{n+1}}\right)^2\right]}{\left(\varepsilon 2^{-b}(\tau - t_{n+1})^{-b}\right)^2} \leq \frac{4^{b+1}C\left(\|c\|_\infty\right)^2}{\varepsilon^2}(\tau - t_{n+1})^{2b}\int_0^{t_{n+1}} (\tau - s)^{\min\{b,1\}-2b-1}\mathrm{d}s. \tag{158}$$

Finally, we have

$$(\tau - t)^{2b} \int_0^t (\tau - s)^{\min\{b,1\}-2b-1}\mathrm{d}s \leq C'(\tau - t)^{\min\{b,1\}},\quad 0 \leq t < \tau \tag{159}$$

with some $C' > 0$. Then, we obtain

$$\sum_{n=1}^{+\infty} \mathbb{P}(A_n) \le \frac{4^{b+1} CC' \left(\|c\|_\infty\right)^2}{\varepsilon^2} \sum_{n=1}^{+\infty} (\tau - t_{n+1})^{\min\{b,1\}} = \frac{4^{b+1} CC' \left(\|c\|_\infty\right)^2 \tau^{\min\{b,1\}}}{\varepsilon^2} \sum_{n=1}^{+\infty} \frac{1}{2^{(n+1)\min\{b,1\}}} < +\infty . \quad (160)$$

By Borel–Cantelli lemma (Theorem 2.7 in Klenke [84]), we obtain $\mathbb{P}\left( \limsup_{n \to +\infty} A_n \right) = 0$. This combined with the continuity of path of $z_t$ ($t < \tau$) yields $z_t \to 0$ with probability 1 because at most a finite number of $A_n$ occur and $\varepsilon > 0$ is arbitrary.

## A3. Supporting data

**Table A1** lists the primers and probes used in this study. **Table A2** summarizes the PCR conditions in this study. **Table A3** shows the observed eDNA data at each site. **Figure A1** compares the temporal profiles of the average (Ave) of eDNA concentrations for Kisuki at the left-most grid point and the theoretical result at $x = 0$. **Figure A2** shows the results for Shin-Mitoya. As shown in these figures, the finiteness of the domain length $L$ mainly affects the tail of the concentration at $x = 0$ for the Kisuki case.

**Table A1.** List of primers and probes used in this study.

| Primer | Probes (5'→3') |
|---|---|
| Paa-CyB-Forward | CCTAGTCTCCCTGGCTTTATTCTCT |
| Paa-CyB-Reverse | GTAGAATGGCGTAGGCGAAAA |
| Paa-CyB-Probe | FAM-ACTTCACGGCAGCCAACCCCC-TAMRA |

**Table A2.** PCR conditions.

| Step | Temperature (°C) | Duration (s) | No. of cycles | No. of iterations |
|---|---|---|---|---|
| Early-stage denaturation | 95 | 60 | - | |
| Denaturation | 95 | 15 | 50 | 3 |
| Annealing | 62 | 30 | | |

**Table A3.** Observed eDNA concentrations (copies/ml) at each site in 2025.

| Date | Kisuki | Shin-Mitoya |
|---|---|---|
| April 3 | 0* | 0* |
| April 9 | 0* | 0* |
| April 17 | 7.73 | 3.12 |
| April 24 | 4.71 | 9.80 |
| April 30 | 13.31 | 39.51 |
| May 8 | 18.39 | 105.28 |
| May 14 | 48.65 | 314.35 |
| May 21 | 165.49 | 459.80 |
| May 28 | 135.11 | 512.89 |
| June 5 | 171.22 | 494.60 |
| June 12 | 271.13 | 463.38 |
| June 19 | 117.45 | 81.44 |
| July 2 | 56.83 | 131.72 |
| July 9 | 48.05 | 144.56 |

* Not detected. We understood this by the concentration value 0.

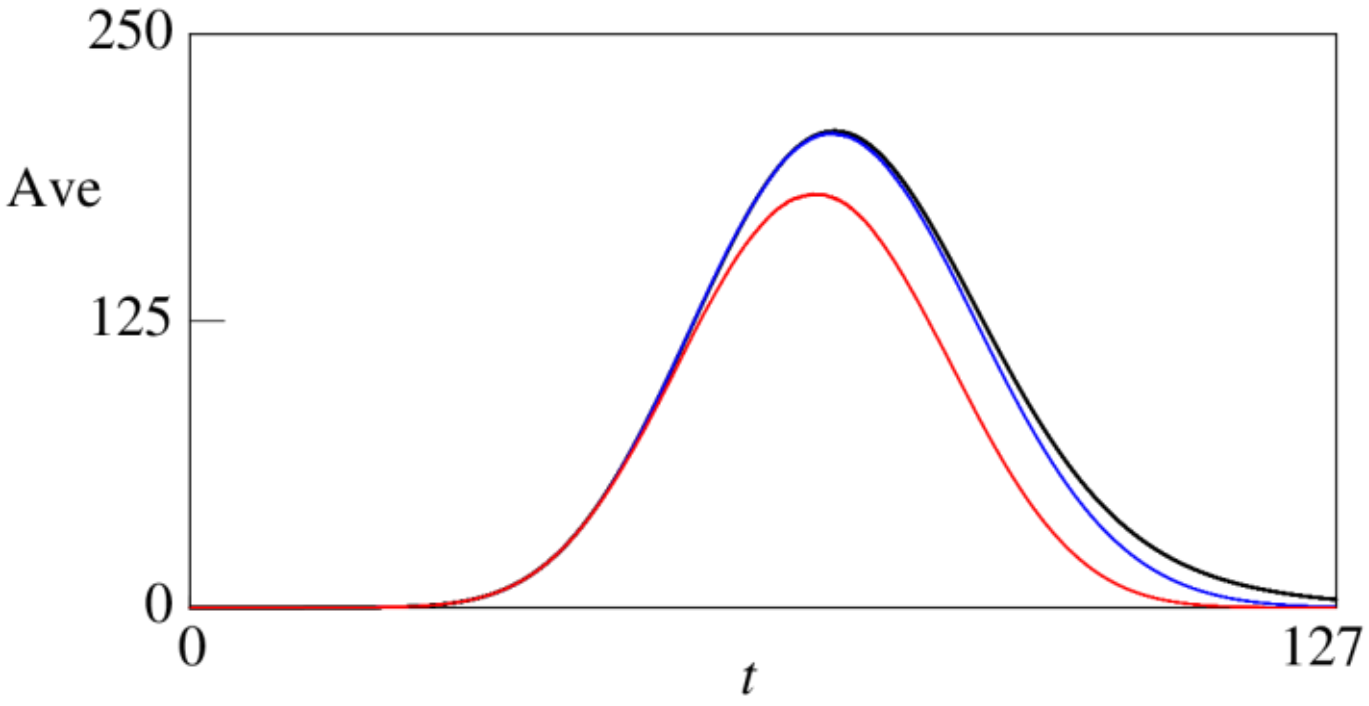

**Figure A1.** Profiles of the average (Ave) of eDNA concentrations $Y$ (copies/ml) for Kisuki at the left-most grid point (blue: $L = 30$ (km), red: $L = 15$ (km)) and the theoretical result at $x = 0$ (black).

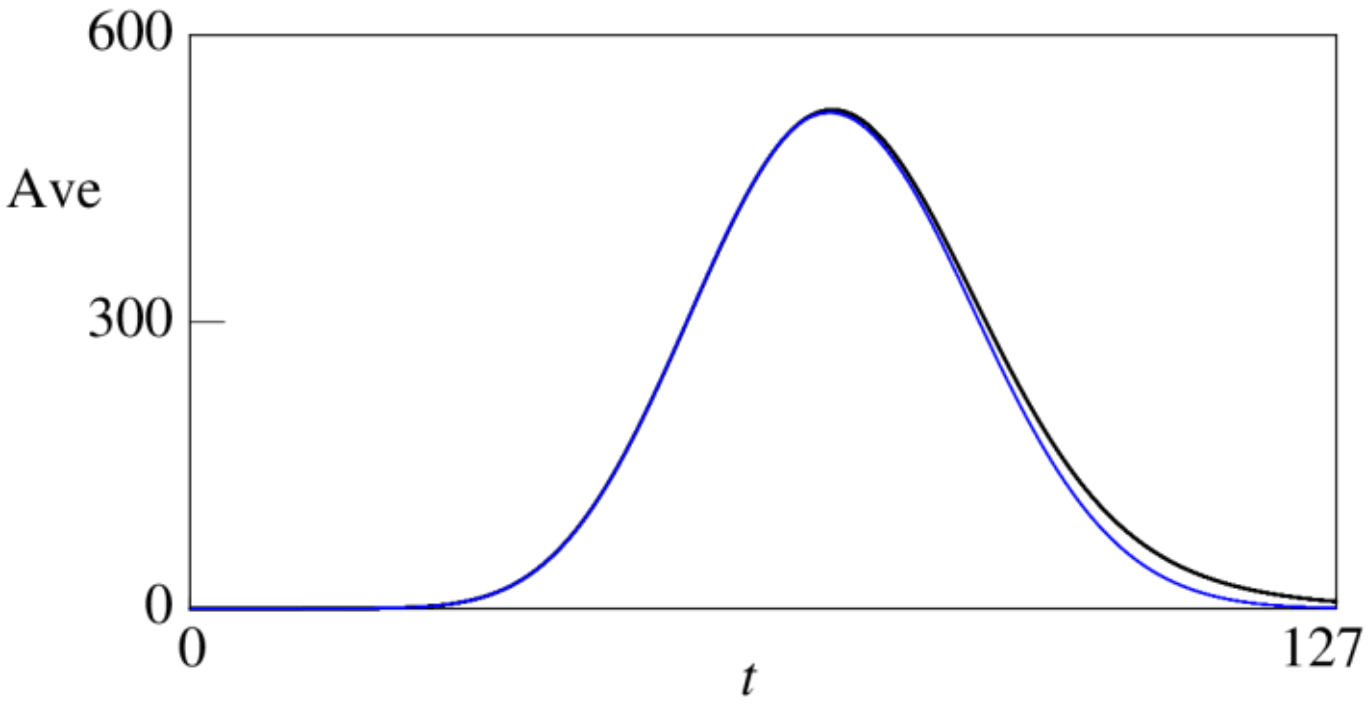

**Figure A2.** Profiles of the average (Ave) of eDNA concentrations $Y$ (copies/ml) for Shin-Mitoya at the left-most grid point (blue) and the theoretical result at $x = 0$ (black).